\documentclass[reprint,nofootinbib,amsmath,amssymb,fontenc aps]{revtex4-1}
\usepackage{graphicx,dcolumn,bm,hyperref,float}
\usepackage{anyfontsize}
\usepackage{color}

\newcommand{\be}{\begin{equation}}
\newcommand{\ee}{\end{equation}}
\newcommand{\beq}{\begin{equation}}
\newcommand{\eeq}{\end{equation}}
\newcommand{\ba}{\begin{eqnarray}}
\newcommand{\ea}{\end{eqnarray}}
\newcommand{\bef}{\begin{figure}}
\newcommand{\eef}{\end{figure}}

\newcommand{\m}{m_\phi}
\newcommand{\Om}{\Omega_\phi}
\newcommand{\Td}{T_d}
\newcommand{\sig}{\sigma_\phi}
\newcommand{\mpl}{M_{\mbox{\tiny{Pl}}}}
\newcommand{\Mmax}{M_{\mbox{\tiny{max}}}}
\newcommand{\Imax}{I_{\mbox{\tiny{max}}}}
\newcommand{\Ms}{M_{\text{\tiny{sun}}}}
\newcommand{\Lamin}{\Lambda_{\mbox{\tiny{bc}}}}
\newcommand{\A}{\bar{A}}
\newcommand{\B}{\bar{B}}
\newcommand{\lambdaeff}{\lambda_{4\mbox{\tiny{eff}}}}
\newcommand{\Menc}{M_{\mbox{\tiny{enc}}}}

\newcommand{\bp}{\beta}
\newcommand{\F}{g}
\newcommand{\en}{\rho}

\newcommand{\GamD}{\Gamma_{\mbox{\tiny{d}}}}
\newcommand{\pref}{\alpha}
\newcommand{\prefC}{\beta}
\newcommand{\prefmu}{\gamma}
\newcommand{\prefmuC}{\delta}
\newcommand{\prefD}{\epsilon}
\newcommand{\prefmuD}{\eta}
\newcommand{\C}{C}
\newcommand{\Rs}{R_S}
\newcommand{\Rp}{R_{*}}
\newcommand{\bc}{\zeta_u}
\newcommand{\cc}{\zeta_l}

\newcommand{\phia}{\phi_a}
\newcommand{\omegah}{\omega_0}

\newcommand{\cs}{\chi}
\newcommand{\N}{P}
\newcommand{\GamC}{\Gamma_{\mbox{\tiny{class}}}}
\newcommand{\lambdaft}{\lambda_{\mbox{\tiny{42}}}}
\newcommand{\lambdatt}{\lambda_{\mbox{\tiny{32}}}}

\begin{document}

\title{Decay of Boson Stars with Application to Glueballs and Other Real Scalars}

\author{Mark P.~Hertzberg}
\email{mark.hertzberg@tufts.edu}
\author{Fabrizio Rompineve}
\email{fabrizio.rompineve@tufts.edu}
\author{Jessie Yang}
\email{jessie.yang@tufts.edu}
\affiliation{Institute of Cosmology, Department of Physics and Astronomy, Tufts University, Medford, MA 02155, USA
\looseness=-1}

\begin{abstract}
One of the most interesting candidates for dark matter are massive real scalar particles. A well-motivated example is from a pure Yang-Mills hidden sector, which locks up into glueballs in the early universe. The lightest glueball states are scalar particles and can act as a form of bosonic dark matter. If self-interactions are repulsive this can potentially lead to very massive boson stars, where the inward gravitational force is balanced by the repulsive self-interaction. This can also arise from elementary real scalars with a regular potential. In the literature it has been claimed that this allows for astrophysically significant boson stars with high compactness, which could undergo binary mergers and generate detectable gravitational waves. Here we show that previous analyses did not take into proper account $3\to2$ and $4\to2$ quantum mechanical annihilation processes in the core of the star, while other work misinterpreted the classical $3\to1$ process. In this work, we compute the annihilation rates, finding that massive stars will rapidly decay from the $3\to2$ or $4\to2$ processes (while the $3\to1$ process is typically small). Using the Einstein-Klein-Gordon equations, we also estimate the binding energy of these stars, showing that even the densest stars do not have quite enough binding energy to prevent annihilations. For such boson stars to live for the current age of the universe and to be consistent with bounds on dark matter scattering in galaxies, we find the following upper bound on their mass for $\mathcal{O}(1)$ self-interaction couplings: $M_* \lesssim 10^{-18} \Ms$ when $3\to2$ processes are allowed and $M_* \lesssim 10^{-11} \Ms$ when only $4\to2$ processes are allowed. We also estimate destabilization from parametric resonance which can considerably constrain the phase space further. Furthermore, such stars are required to have very small compactness to be long lived.
\end{abstract}

\maketitle

\tableofcontents

\section{Introduction}

Perhaps the best motivation for physics beyond the Standard Model is the presence of dark matter which comprises most of the mass of the universe. The most exciting aspect of dark matter is that it may represent an entirely new sector of physics. Due to the current lack of discovery of any dark matter candidates that have direct couplings to the Standard Model, including WIMPs, it raises the possibility that dark matter may be part of some hidden sector and/or associated with new very heavy particles (e.g., see Refs.~\cite{Kaplan:2009ag,Cohen:2010kn,Shelton:2010ta,Cheung:2010gj,Das:2010ts,Foot:2014uba,Blinov:2012hq,Boddy:2014yra,Lonsdale:2014wwa,Buckley:2014fba,Elor:2015bho,Acharya:2016fge,Dienes:2016vei,Escudero:2017yia,Tsao:2017vtn,Hertzberg:2019bvt}). 

Hidden sectors may include pure Yang-Mills interactions, i.e., new collections of massless spin 1 particles with self interactions. Such constructions are entirely plausible from the point of view of fundamental physics. They are also considered to be entirely natural, since they do not appeal to any unnecessarily small parameters. In particular, they have no allowed elementary masses, and are fully described by only two quantities: the scale at which the theory becomes strong coupled $\Lambda$ (which can be naturally small compared to any unification scale due to the logarithmically slow running of coupling) and the size of the gauge group, such as $SU(N)$. It is also possible of course that physics beyond the Standard Model includes various other kinds of particles. Importantly, this may include elementary spin 0 particles. Both the hidden spin 1 and spin 0 particles are of course bosons, and as such they allow for a rich phenomenology; under some conditions they can organize into interesting states of high occupancy, as we will discuss in this paper.

Let us address in more detail the case of hidden Yang-Mills. If there are indeed such sectors of interacting spin 1 particles, it leads to various questions: What happens below the confinement scale? If the particles are thermally produced in the early universe, what is their energy density today? Do the strong interactions lead to inconsistency with constraints on dark matter scattering in galaxies? And, very importantly, are there novel observational signatures of these sectors? 

To address these issues, let us start in the very early universe. We know that at high temperatures of the early universe, the particles exhibit asymptotic freedom, and so we have a gas of almost free particles. When the hidden sector temperature is on order of the strong coupling scale $\sim\Lambda$ the particles are expected to lock up into color neutral states, known as ``glueballs" (for a review, see Ref.~\cite{Mathieu:2008me}). The lightest glueball states are expected to be spin 0 particles, whose mass $\m$ is of the order the strong coupling scale $\Lambda$. As the temperature of the dark sector $\Td$ lowers further, these glueballs become non-relativistic, and can act as a form of dark matter in the late universe. If one just focusses on freeze-out, the relic abundance can be estimated to be roughly \cite{Boddy:2014yra} 
$\Om\sim N^2\,\xi^3(\Lambda/(100\,\mbox{eV}))$ 
where $\xi\equiv \Td/T$ is the ratio of temperatures of the dark to visible sectors. If $\xi\sim1$ then one evidently needs rather small values of $\Lambda$, no larger than a few eV, to avoid overproduction. However, there are important regimes in which the abundance can be modified due to self-interactions leading to $3\to2$ processes (this will be relevant to the rates we compute in stars later). For aspects of glueball and self-interacting dark matter relic abundances, see Refs.~\cite{Carlson:1992fn,Hochberg:2014dra,Bernal:2015ova,Hochberg:2015vrg,Pappadopulo:2016pkp,Forestell:2016qhc,Farina:2016llk,Halverson:2016nfq}.

On the other hand, if the temperature of the hidden sector is small ($\xi\ll1$) then one can have larger strong coupling scales without overclosing the universe. As some of us showed in Ref.~\cite{Hertzberg:2019prp}, there is a very reasonable scenario in which the inflaton $\varphi$ decays predominantly to the Standard Model through the dimension 3 coupling to the Higgs $\varphi\, H^\dagger H$, while its decays to hidden sector Yang-Mills is suppressed as it would  occur through the dimension 5 coupling $\varphi\, G_{\mu\nu}\,G^{\mu\nu}$. This leads to the expectation $\xi\ll 1$ (such as $\xi\approx 0.006$ for reasonable parameters considered in Ref.~\cite{Hertzberg:2019prp}). 

Furthermore, glueballs exhibit scattering in galaxies. The scattering cross section to mass ratio is expected to be on the order $\sig/\m\sim \mbox{few}/(\Lambda^3 N^4)$ \cite{Soni:2016gzf}. While constraints from bullet cluster implies $\sig/\m\lesssim$\,\,cm$^2/$gram\,\,$\approx 1/(60\,\mbox{MeV})^3$ \cite{Markevitch:2003at,Harvey:2015hha}. 
Thus to satisfy this bound, one needs $\Lambda\gtrsim\Lamin$ where $\Lamin\sim 100\,\mbox{MeV}/N^{4/3}$. This in turn implies that $\Om$ is large unless $\xi\ll 1$, which is compatible with the reasoning in Ref.~\cite{Hertzberg:2019prp} (although for extremely high $N$, one would need to compensate with extremely small $\xi$, which seems less plausible). Or alternatively, if $3\to2$ processes are significant, this can reduce the abundance further. 

If we move beyond the glueball motivation, we can in fact consider any dark matter candidate which organizes into massive scalars with self-interactions. As an example, we could just study an elementary scalar with a renormalizable potential. Generally, these are all interesting candidates that are the focus of this study. 

\subsection{Novel Signature}

An interesting possibility is that the dark matter scalars organize into gravitationally bound systems, so-called ``boson stars" (for a review see Refs.~\cite{Jetzer:1991jr,Schunck:2003kk,Chavanis:2011cz,Liebling:2012fv}). For both glueballs and for elementary scalars with a regular potential, one anticipates self-interactions, including $\sim\lambda_4\phi^4$, etc. For repulsive self-interactions this can give rise to very dense boson stars. This was carefully studied originally in Ref.~\cite{Colpi:1986ye} for a complex scalar field with a global $U(1)$ symmetry (recent work includes Ref.~\cite{Choi:2019mva}). They solved the full Einstein-Klein-Gordon equations of motion for spherically symmetric time independent solutions, finding that the maximum star mass is 
\beq
\Mmax\sim{\sqrt{\lambda_4}\,\mpl^3\over\m^2}
\label{ColpiMaxmass}\eeq
($\mpl\equiv1/\sqrt{G}\approx1.2\times 10^{19}$\,GeV). 
At this maximum mass, the physical radius is only a factor of $\sim 2$ larger than the corresponding Schwarzschild radius. For masses above $\Mmax$, boson stars solutions do not exist; the density is so high that such configurations can collapse to a black hole. Interestingly, if $\lambda_4=\mathcal{O}(1)$, then this scaling is similar to the Chandrasekhar mass for white dwarf stars $M_{\mbox{\tiny{wd}}}\sim\mpl^3/m_{\mbox{\tiny{p}}}^2$; one is effectively replacing a repulsion from Pauli exclusion of fermions by a self-interaction repulsion of bosons. Then if $\m\lesssim$\,GeV, this can be of the order of a solar mass $\Ms\sim 10^{57}$\,GeV, or larger.

In the literature, this result has been applied to boson stars from real scalars, including glueballs and elementary scalars. It is not clear that glueballs have the required repulsive interaction (and in the case of the $SU(2)$ gauge group, some recent lattice calculations suggest it may in fact be attractive \cite{Yamanaka:2019aeq,Yamanaka:2019yek}). Since we do not know the sign for a generic gauge group, we will simply work under the assumption that it may be repulsive for some cases, and this is of course certainly possible for scalars in other kinds of theories. To apply the result to glueballs, one estimates the quartic coupling as $\lambda_4\sim (4\pi)^2/N^2$ and $\m\sim\mbox{few}\,\Lambda$. Then using the above bound on $\Lambda\gtrsim\Lamin$ in order to satisfy bullet bluster constraints, one finds that the maximum mass can be as large as $\Mmax\sim\mpl^3N^{5/3}/(100\,\mbox{MeV})^2\sim 100 \Ms N^{5/3}$. 

Hence one can readily have boson star masses that are larger than a solar mass by choosing the strong coupling scale accordingly. Alternatively, if one is simply studying elementary scalars, one can just impose the appropriate values of $\m$ and $\lambda_4$ to obtain such massive stars and satisfying bullet cluster constraints.

Very interestingly, Refs.~\cite{Soni:2016gzf,Soni:2016yes,daRocha:2017cxu,Soni:2017nlm} pointed out that if one has $\Mmax\gtrsim\mathcal{O}(10)\Ms$ and if these boson stars undergo a merger, they can emit gravitational waves with a frequency and amplitude that could possibly be detectable at LIGO, and even heavier stars at LISA, with a signal that could be potentially distinguished from that of black hole mergers (for a focus on complex scalars, see Refs.~\cite{Croon:2018ybs,Guo:2019sns}). This presents a novel signature of these hidden sectors. This is such an exciting possibility, it acted as a motivation for the present work. 


\subsection{Outline of this work}

In this work, we critically examine whether such stars are in fact long lived. While a complex scalar with an internal $U(1)$ symmetry, of the sort studied in Ref.~\cite{Colpi:1986ye}, carries a conserved particle number, the same is not true for a real scalar. In this case there is no distinction between the particle and its antiparticle. Such real scalars can undergo quantum annihilation processes in the core of the star, including $3\phi\to2\phi$ and $4\phi\to2\phi$ processes. Such processes will cause the star to evaporate away. In this paper we compute these annihilation rates for the boson stars, finding that while the processes are negligible for very low mass stars, they are extremely important for high mass stars including those above a solar mass. We find that this leads to short lifetimes, unless the couplings are very small (associated with huge gauge groups for glueballs). However, even for small couplings, the stars whose mass would be relevant to LIGO/LISA typically have low compactness and so the gravitational wave emission would be suppressed (unless one takes extreme parameters). Furthermore, we estimate possible decays from parametric resonance, finding that this cuts down the available parameter space considerably further. 

One might hope that the star carries enough binding energy to prevent these number changing processes. We find that even though the most massive stars have appreciable binding energy, it is not sufficient to prevent the radiation. Finally, for completeness, we also compute $3\phi\to1\phi$ processes, which had previously been claimed to be the most important process in the work of Ref.~\cite{Eby:2015hyx} (in the context of axions). We show that the calculations provided in that work, while appearing as a quantum mechanical process, are in fact properly captured by classical field theory when the re-scaling to the final decay rate of the condensate is obtained, and we discuss the pre-factor. For the massive stars of interest, these $3\phi\to1\phi$ classical processes are found to typically be small, while the $4\phi\to2\phi$ or $3\phi\to2\phi$ quantum processes that we focus on can be very important.

Our paper is organized as follows:
In Section \ref{SFT} we present the basic effective field theory field theory, the corresponding classical equations of motion for a spherically symmetric star, and recap the properties of stars in a single harmonic approximation.
In Section \ref{Annihilation} we compute the quantum annihilation rates in the core of the stars.
In Section \ref{Bounds} we use these results to derive bounds on the mass and compactness. 
In Section \ref{Resonance} we also consider parametric resonance.
In Section \ref{Binding} we estimate whether the star's binding energy can prevent decays. 
In Section \ref{Classical} we compute the classical radiation and compare to existing claims in the literature. 
In Section \ref{PressureStars} we briefly discuss decays in boson stars supported by quantum pressure.
Finally, in Section \ref{Conclusions} we conclude.

\vspace{-0.1cm}

\section{Boson Stars}\label{SFT}

When glueballs form they organize into spin 0 scalar particles. In the effective field theory formalism, we can organize this into a scalar field $\phi$. These scalars interact directly with one another in the effective theory, which is a consequence of the microscopic hidden gluon interactions. In this work, the most important form of the interactions will be the leading order operators as organized into a scalar potential $V$. We can expand this as
\be
V(\phi)={1\over2}\m^2\phi^2+{\lambda_3\,\m\over3!}\phi^3+{\lambda_4\over4!}\phi^4+{\lambda_5\over5!\,\m}\phi^5+\ldots\label{V}
\ee
where $\lambda_n$ are dimensionless couplings. There could be a tower of higher dimension operators, including higher order derivative operators in the effective theory, but these leading terms will be sufficient to describe the most relevant effects in this work. This formalism is obviously also applicable to other scalars, including elementary scalars with a renormalizable potential. So our analysis will be quite general. In the case of glueballs formed from the group $SU(N)$, the couplings are expected to scale as $\lambda_{n}\sim(4\pi/N)^{n-2}$, with $n=3,4,\ldots$. For a generic real scalar the same hierarchy of couplings may occur too
\be
\lambda_{n}\sim\lambda^{(n-2)/2}\,\,\,\,\,\,\,\,\,(\lambda\equiv\lambda_4)
\ee
But we do not suppose that the prefactors are necessarily close to 1. In the case of a real scalar, one can also imagine that it is endowed with some discrete $\mathcal{Z}_2$ symmetry $\phi\to-\phi$, forbidding all the odd powers. We shall consider this case in this work also. 

Of course the field is also coupled to gravity with action (we use units $c=\hbar=1$ and signature $-+++$) 
\be
S=\int d^4x\sqrt{-g}\left[{\mathcal{R}\over 16\pi G}-{1\over2}g^{\mu\nu}\partial_\mu\phi\partial_\nu\phi-V(\phi)\right]
\label{Saction}
\ee
where $\mathcal{R}$ is the Ricci scalar, $G$ is Newton's gravitational constant, and $g_{\mu\nu}$ is the metric of space-time.

\subsection{Spherical Symmetry}

Under some conditions, the scalars can condense into states of high occupancy, which is usually well described by classical field theory (we shall return to quantum corrections in the next section). A condensate for a system with gravity is a localized system; either a gravitationally bound ``boson star", or a potentially bound (for attractive interactions) ``oscillon". This work will focus on repulsive interactions mainly (though some of our reasoning will be relevant for attractive interactions too), so the only relevant solution is the gravitationally bound boson star. The lowest energy configurations are expected to be spherically symmetric.

In this case, it is useful to go to spherical co-ordinates $(r,\theta,\varphi)$, where quantities only depend on radius $r$ (as well as time $t$). Any spherically symmetric space-time metric can be written in the Schwarzschild co-ordinates
\be
ds^2=-B(r,t)dt^2+A(r,t)dr^2+r^2(d\theta^2+\sin^2\!\theta\, d\varphi^2)
\ee
where $g_{rr}=A$ and $g_{tt}=B$ are functions that can depend on radius and time that we need to solve for; we also need to solve for the scalar field $\phi(r,t)$.

\subsection{Einstein-Klein-Gordon Equations}

The Einstein equations $G_{\mu\nu}=8\pi G\, T_{\mu\nu}$ follow from varying the action $S$ above with respect to $g_{\mu\nu}$ and extremizing. Here $G_{\mu\nu}$ is the Einstein tensor and $T_{\mu\nu}$ is the energy-momentum tensor provided by the scalar field $\phi$. 
Due to spherical symmetry, we only need to report on the time and radius components of the Einstein equations. The Einstein tensor can be shown to be
\ba
&&G_{tt}=B\left({A'\over r\,A^2}+{1\over r^2}\left(1-{1\over A}\right)\right)\label{Gtt}\\
&&G_{rr}=A\left({B'\over r\,A\,B}-{1\over r^2}\left(1-{1\over A}\right)\right)\label{Grr}
\ea
Here we use a notation in which a prime means a derivative with respect to radius $r$ and a dot means a derivative with respect to time $t$.

The energy momentum tensor arises from varying the scalar field contribution of $S$ with respect to the metric
\be
T_{\mu\nu}=\partial_\mu\phi\partial_\nu\phi + g_{\mu\nu}\mathcal{L}_\phi
\ee
The time and radius components are readily found to be
\ba
&&T_{tt}={1\over2}\dot\phi^2+{B\over 2A}(\phi')^2+B\, V\label{Ttt}\\
&&T_{rr}={A\over2B}\dot\phi^2+{1\over 2}(\phi')^2-A\, V\label{Trr}
\ea
Combining eqs.~(\ref{Gtt},\ref{Grr}) with eqs.~(\ref{Ttt},\ref{Trr}) we have a pair of Einstein equations for $A$ and $B$. Note that these equations only involve spatial derivatives for $A$ and $B$, which reflects the fact that these are constrained variables; there are no dynamical (tensor) components of the metric due to spherical symmetry.

Finally we need the equation of motion for the scalar field $\phi$. This comes from varying the action $S$ with respect to $\phi$, giving 
\be
\partial_\mu\left(\sqrt{-g}\,g^{\mu\nu}\partial_\mu\phi\right)=\sqrt{-g}\,V'
\ee
Substituting in the above metric and carrying out the derivatives leads to
\be
\phi''+\left[{2\over r}+{B'\over 2B}-{A'\over 2A}\right]\!\phi'-{A\over B}\left[\ddot\phi+\left({\dot{A}\over 2A}-{\dot{B}\over 2B}\right)\dot\phi\right] = A\,V' \label{KG}
\ee
There is also a 3rd Einstein equation from the time-space components, which arises from $G_{tr}=\dot{A}/(r A)$ and $T_{tr}=\dot\phi\,\phi'$. However, this is redundant with the above set of equations and so is not needed.

\subsection{Single Harmonic Approximation}\label{SHA}

In general a solution for a real scalar field will have a complicated time dependence. A boson star $\phi_*$ or oscillon is a periodic solution of the classical equations of motion. It can therefore be expanded in a harmonic expansion
\be
\phi_*(r,t)=\sum_{n=1} \Phi_n(r)\cos(\omega_n t)
\ee
(this solution exists up to exponentially small corrections, which we will return to in Section \ref{Classical}).
For small amplitude stars one can be sure that this expansion is dominated by the first harmonic with $\omega$ close to, but slightly less than, the mass $\m$. For very large amplitude stars, especially those that are at masses approaching the maximum mass, higher order terms can become more important. By inserting this into the above equations of motion, along with a similar harmonic expansion for metric components $A(r,t)$ and $B(r,t)$, one obtains an infinite tower of coupled non-linear ordinary differential equations for the prefactors $\Phi_n(r),\,A_n(r),\,B_n(r)$. In principle, one can numerically solve this complicated system. Interesting work on this was done in Refs.~\cite{ValdezAlvarado:2011dd,Mahmoodzadeh:2017dpp} for the $\phi^4$ potential, where the tower of harmonics was constructed and numerical investigations were done. However, we leave a full treatment of all this for future work. 

A useful approximation is to assume that the system is dominated by only one harmonic of frequency $\omega$. We write
\be
\phi_*(r,t)\approx\sqrt{2}\,\Phi(r)\cos(\omega t)
\ee
(the factor of $\sqrt{2}$ is for convenience).
This is accurate at small amplitudes, as it is dominated by the fundamental in that case anyhow, and provides a rough estimate of the behavior at high amplitudes too. Of course by inserting this into the equations of motion it will not satisfy them precisely as it will generate higher harmonics. So the idea of the single harmonic approximation is to time average the equations of motion over a period of oscillation $T=2\pi/\omega$. Self-consistently then, we need to approximate the metric as time independent, writing $A(r,t)\approx\A(r)$ and $B(r,t)\approx\B(r)$. On the right hand side of the Einstein equations we need the time average of the energy momentum tensor, which is
\ba
&&\langle T_{tt}\rangle={1\over2}\omega^2\Phi^2+{B\over 2A}(\Phi')^2+B\langle V\rangle\label{TttAv}\\
&&\langle T_{rr}\rangle={A\over2B}\omega^2\Phi^2+{1\over 2}(\Phi')^2-A\langle V\rangle\label{TrrAv}
\ea
In addition, a useful way to handle the Klein-Gordon equation in eq.~(\ref{KG}) is to multiply throughout by $\sqrt{2}\cos(\omega t)$ and then time average. This gives
\be
\Phi''+\left[{2\over r}+{B'\over 2B}-{A'\over 2A}\right]\!\Phi'+{A\over B}\,\omega^2\,\Phi = \sqrt{2}\,A\left\langle\cos(\omega t)V'\right\rangle
\ee

The time average of the potential $V$ is simple if the potential has only even powers of $\phi$. If there are odd powers of $\phi$, like $\sim\lambda_3\m\phi^3$, then the averaging is more subtle. Naively one just time averages those terms to zero. But in fact the presence of those terms means that the field undergoes asymmetric oscillations; it spends more time on one side of the potential than the other. A proper treatment of this would include a time independent term in $\phi$, providing a non-zero mean $\langle\phi\rangle\neq 0$ (e.g., see the asymmetric terms in the study of oscillons in Ref.~\cite{Fodor:2008du}). 

However, we can instead pass to the low momentum effective theory; this is valid since we will only be studying boson star solutions whose size is much larger than the inverse mass of the scalar $\m$. Here one integrates out intermediate $\phi$ exchange processes from $\sim\lambda_3\m\phi^3$. This replaces interactions by only contact $\sim\lambda\phi^4$ interactions, i.e., at large distances the interaction between scalars acts a kind of delta-function interaction. In the low momentum effective theory this leads simply to a shift in the effective quartic coupling, which one can readily show is (e.g., see Ref.~\cite{Hertzberg:2010yz})
\be
\lambdaeff=\lambda_4-{5\over3}\lambda_3^2
\ee
In the effective theory one needs $\lambdaeff>0$ in order to have the repulsive interaction needed to support the stars of interest in this work (stars supported by quantum pressure are discussed briefly in Section \ref{PressureStars}).
The corresponding time averages are then (we truncate the potential to quartic order here for simplicity)
\ba
\langle V\rangle &=& {1\over2}\m^2\Phi^2+{\lambdaeff\over16}\Phi^4\\
\sqrt{2}\langle\cos(\omega t)V'\rangle &=& \m^2\Phi+{\lambdaeff\over4}\Phi^3
\ea

Bound state solutions are found by numerically searching for configurations that obey the boundary conditions $\A\to1,\B\to1,\,\Phi\to0$ as $r\to\infty$ and $\Phi'\to0$ as $r\to0$. The ground state solution is identified by having no nodes.

One can readily integrate up the $\langle G_{tt}\rangle=8\pi G \langle T_{tt}\rangle$ equation to solve for $\A$ in terms of the enclosed mass $\Menc(r)$ as
\be
\A(r) = \left(1-{2G\Menc(r)\over r}\right)^{\!-1}
\ee
The enclosed mass is just defined as the (weighted) integral of the energy density
\be
\Menc(r) = 4\pi\int_0^r\,dr'\,r'^2\,\langle T_{tt}(r')\rangle/\B(r')
\ee
The corresponding mass of the star, or total energy, is 
\be
M_* = \Menc(\infty)
\ee

Note that $\A\to1$ as $r\to0$. However, the value of $\B$ or $\Phi$ as $r\to0$ is not specified uniquely. Let us call them $\B_0\equiv\B(0)$ and $\Phi_0\equiv\Phi(0)$, respectively. Their values are related to the value of $\omega$. It is useful to trade in $\B$ for another dimensionless function as
\be
\bp(r)\equiv{\omega^2\over\m^2\,\B(r)}
\ee
By scanning different values of $\bp_0\equiv\bp(0)$, or equivalently by scanning different values for $\Phi_0$, and numerically solving the equations, one finds a range of boson star solutions. An example is given in Fig.~\ref{StarSolution} (orange curve) for the choice $\Phi_0\approx0.028\,\mpl$, where we have defined the Planck mass $\mpl\equiv1/\sqrt{G}$.

\subsection{Large Coupling Regime}\label{LCR}

As shown in Ref.~\cite{Colpi:1986ye}, the structure of the solutions is controlled by the following dimensionless parameter
\be
\F\equiv {\lambdaeff\mpl^2\over16\pi\,\m^2}
\ee
(Note: our $\F$ is what Ref.~\cite{Colpi:1986ye} calls $\Lambda$, but we have used $\Lambda$ as strong coupling scale,
and our $\lambdaeff/4$ is effectively playing the role of $\lambda$ in Ref.~\cite{Colpi:1986ye}.)
If $\F$ is small, then the self-interactions are negligible and the solution is just provided by gravity balanced by quantum pressure of the bosons (which is classical pressure within the field formalism); we shall study this in Section \ref{PressureStars}.  However, if $\F$ is large, then the behavior changes: there exist massive solutions in which gravity is balanced by the repulsive self-interaction, which will be the main focus of this work (in addition there always exists very light solutions in which one can once again ignore self-interactions, but they are less interesting to us here). For glueballs, one anticipates $\m\lll\mpl$ and $\lambdaeff$ a parameter that is not especially small, with $\lambdaeff\sim (4\pi/N)^2$. If we take $N=\mathcal{O}(2)$ and $\m\sim 0.1$\,GeV then $\F\sim 10^{40}$, i.e., it is extremely large.

\begin{figure}
    \centering
    \includegraphics[width=\columnwidth]{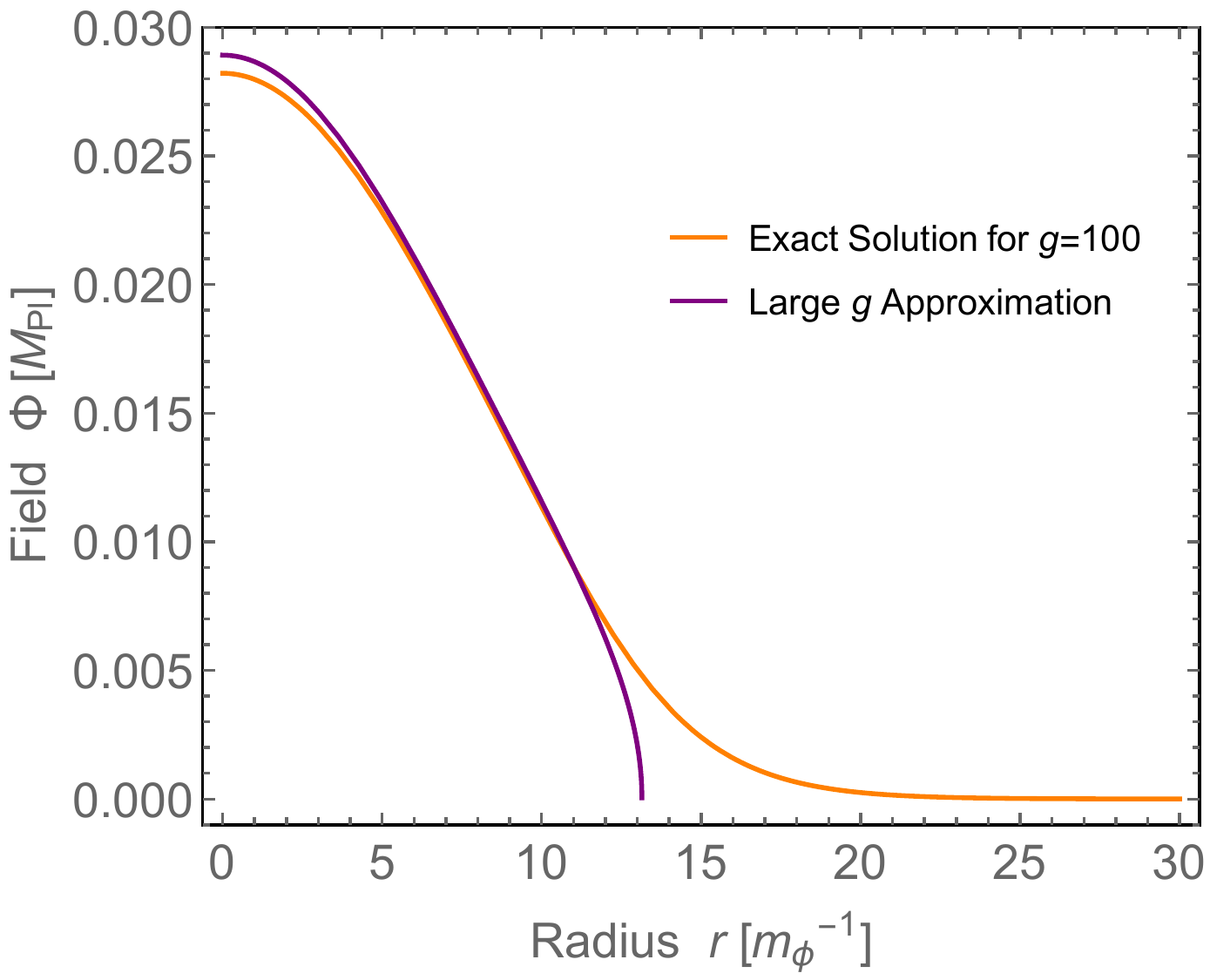}
    \caption{{\em Exact versus large $\F$ approximation profile.}
   Boson star profile $\Phi(r)$ for the core value $\Phi_0\approx0.028\,\mpl$ and parameter $\F=100$. The orange is the exact numerical solution, while the purple is from scaling the large $\F$ approximation (this is similar in form to a plot in Ref.~\cite{Colpi:1986ye}.)}
    \label{StarSolution}
\end{figure}

In this large $\F$ regime, one can further simplify the equations of motion (as pointed out in Ref.~\cite{Colpi:1986ye}).
To make it manifest which terms are important and which terms can be ignored at large $\F$, it is useful to pass to the dimensionless variables
\be
\hat{r}\equiv \m\,r/\sqrt{\F},\,\,\,\hat\Phi\equiv\Phi\sqrt{4\pi\F}/\mpl
\ee
By re-writing the above Einstein-Klein-Gordon equations in terms of these variables and re-scaled metric component $\bp$ and then neglecting all terms that are sub-dominant in the $\F\to\infty$ limit, one obtains the following simplified form of the equations
\ba
\frac{\A^{\hat{'}}}{\hat{r}\,\A^2}+\frac{1}{\hat{r}^2}\left(1-\frac{1}{\A}\right) & = & (\bp+1)\hat{\Phi}^2+{1\over2}\hat{\Phi}^4 \label{approxA}\\
-\frac{\bp'}{\hat{r}\,\A\,\bp}-\frac{1}{\hat{r}^2}\left(1-\frac{1}{\A}\right) & = & (\bp-1)\hat{\Phi}^2-{1\over2}\hat{\Phi}^4 \label{approxB}\\
\bp\,\hat{\Phi}&=&\hat{\Phi}+\hat{\Phi}^3\label{approxsigma}
\ea
One can readily integrate up the equation for $\A$, as above, to obtain the enclosed mass in this limit as
\be
\Menc(\hat{r}) = {\sqrt{\lambdaeff}\,\mpl^3\over4\sqrt{\pi}\,\m^2}\,I(\hat{r})
\label{Mrescale}\ee
where
\be
I(\hat{r})=\int_0^{\hat{r}} d\hat{r}'\,\hat{r}'^{2}\!\left({1\over2}(\bp+1)\hat{\Phi}^2+{1\over4}\hat{\Phi}^4\right)
\ee
In these dimensionless variables, one can anticipate that the maximum value of the integral here, when integrating over the whole star, is $\Imax=\mathcal{O}(1)$ (in addition to a family of lighter stars with $I\ll1$). In fact a precise calculation reveals that $\Imax\approx0.2$.
Thus when multiplying by the prefactor in eq.~(\ref{Mrescale}) gives the maximum star mass $\Mmax\approx0.03\sqrt{\lambdaeff}\,\mpl^3/\m^2$ (see ahead to Fig.~\ref{MassNumberBeta}), consistent with our earlier discussion in the introduction. The corresponding radius of the star is $\Rp\sim\mbox{few}\times G\,\Mmax$, i.e., it is comparable to, though a little larger than, the Schwarzschild radius.

\begin{figure}[t]
    \centering
     \includegraphics[width=0.99\columnwidth]{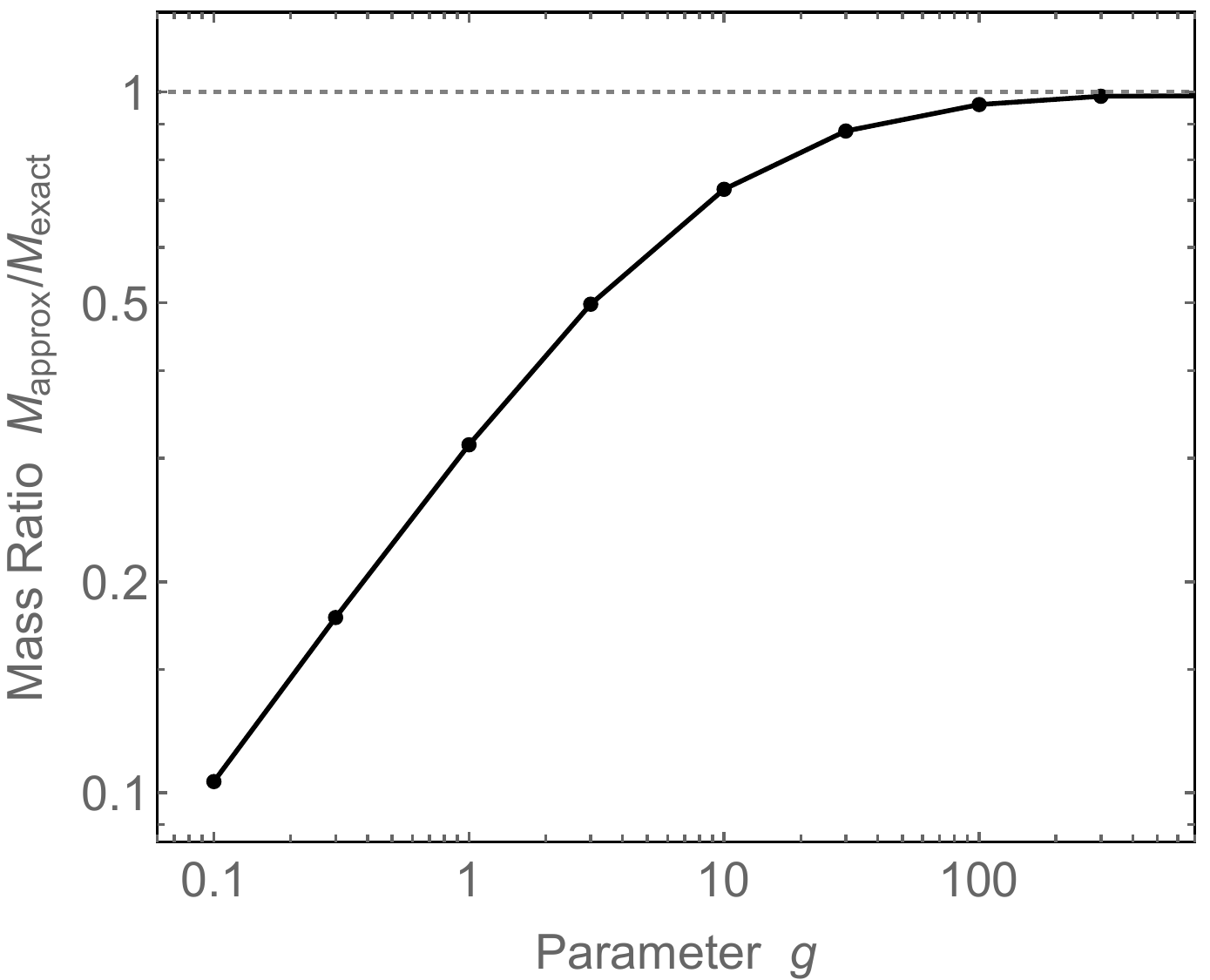}
    \caption{{\em Exact versus large $\F$ approximation mass.}
 Ratio of the mass of a boson star using the large $\F$ approximation scheme to the exact mass using the full equations (albeit always in the single harmonic approximation) with $\bp_0=1.585$. Note that the ratio approaches 1 at large $\F$.}
    \label{StarSolutionMass}
\end{figure}

The Klein-Gordon equation (\ref{approxsigma}) can now be trivially solved $\hat{\Phi}(r)=\sqrt{\bp(r)-1}$. This evidently only makes sense in the regime $\bp>1$. For $\bp<1$ the above approximations break down. This is because even though $\F$ is large the validity of this scaling rests upon the assumption that $\hat{\Phi}$ remains appreciable, say $\mathcal{O}(1)$. However, at large radius the exact solution has an exponentially small tail, as seen in Fig.~\ref{StarSolution} (orange curve), and so this assumption no longer holds. Instead this approximation is accurate in the bulk of the star. One can easily solve these approximate equations to compare. Fig.~\ref{StarSolution} (purple curve) shows a plot for $\F=100$, where the approximation is seen to be somewhat accurate in the bulk. However, as mentioned above, we are mainly interested in extremely large values of $\F$, in which the approximation becomes extremely accurate (away from the exponentially suppressed tail).  

Furthermore, we can compare the star's total mass $M_*$ in the exact versus approximate methods as a function of the dimensionless parameter $\F$. 
We have fixed $\bp_0=1.585$ and numerically obtained the exact equations as well as the approximate result. 
The ratio of the masses is plotted as a function of $\F$ in Fig.~\ref{StarSolutionMass}. Note that the ratio $M_{\mbox{\tiny{approx}}}/M_{\mbox{\tiny{exact}}}\to1$ as $\F$ increases, as expected. For very large $\F$ it becomes prohibitively difficult to solve the exact equations. This is because at fixed $\Phi_0$, one has to search for the corresponding value of $\beta_0$ to obtain a zero node solution that asymptotes to flat space at large radii. This requires continually increasing precision. So instead by passing to the large $\F$ equations we can readily make progress in this regime.

\section{Quantum Annihilation in Cores}\label{Annihilation}

The above analysis is all rigorous in the case of a complex scalar field $\psi$ with a global $U(1)$ symmetry. In that case the above solution is closely related to an exactly harmonic solution by writing $\psi(r,t)=\Phi(r)\,e^{-i\omega t}$. This corresponds to a boson star made out of a collection of particles (or anti-particles). The particles are then stable against annihilation into one another due to the symmetry. One may not expect global symmetries to be exact, but we put aside the discussion of this issue here. For the present purposes of a glueball or single elementary scalar $\phi$, the boson star is made of of $\phi$ particles which are their own anti-particle. Their stability against annihilations is therefore not protected by any symmetry. Since the above very dense stars exist due to the self-interactions $\lambda_n\phi^n$, one should be concerned that these same interactions may lead to rapid annihilation in the star's core. Previous work on this topic has often ignored these details. A notable exception is the work of Ref.~\cite{Eby:2015hyx}, which studied $3\phi\to1\phi$ annihilations (with a focus on axion stars); we shall return to a discussion of that work in Section \ref{Classical}.

\subsection{Perturbative Rates}

To discuss this issue, suppose we had the full classical field boson star solution, including its full time dependence. As we will return to in Section \ref{Classical} this can never be exact due to outgoing classical radiation, but that turns out be typically very small. On the other hand, at any order in the harmonic expansion, one can find a periodic solution that we can refer to as the classical star solution $\phi_*(r,t)$; that will suffice in this section. The approximate classicality can be justified by various considerations, including averaging and decoherence at high occupancy \cite{Hertzberg:2016tal,Allali:2020ttz}. 
In order to discuss the inevitable quantum fluctuations on top of this, we can work in the Heisenberg picture and write the field operator as
\be
\hat{\phi}({\bf x},t) = \phi_*(r,t)+\delta\hat{\phi}({\bf x},t)\label{heisenberg}
\ee
By treating the quantum fluctuations as small $\delta\hat{\phi}\ll\phi_*$, we can write down the Heisenberg equations of motion and linearize them. At small couplings, one can then further obtain solutions for $\delta\hat{\phi}$ working perturbatively. One assumes that the operator begins in the Minkowski vacuum and then evolves the system from there. Since the equations are linear this is doable in principle. If the background $\phi_*$ were homogeneous, the perturbations would be easily diagonalized in $k$-space, leading to standard Floquet theory. However, since our background of interest $\phi_*$ (along with the metric components $A$ and $B$) depend on space, all the $k$-modes are coupled to one another. This makes the analysis non-trivial, even though the system is linear, but can be formulated as a generalized type of Floquet theory.

This was all studied by some of us systematically in Refs.~\cite{Hertzberg:2010yz,Hertzberg:2018zte}; other work appears in Refs.~\cite{
Tkachev:1986tr,Tkachev:1987cd,Tkachev:2014dpa,Kawasaki:2013awa,Mukaida:2016hwd,Eby:2018ufi,Levkov:2020txo}. The important result was that when one is in a small coupling regime, the final resonance matches the result from standard perturbation theory of particles annihilating in vacuum, i.e., the Bose enhancement is shut off for sufficiently small coupling. As shown in that work, the requirement for this is that the maximum parametric resonance Floquet rate $\mu$ from a homogeneous oscillating condensate is smaller than the inverse size of the star $1/\Rp$. The reason for this is that in this regime the resonantly produced scalar particles escape the condensate before Bose-Einstein statistics can be effective. The regime of parametric resonance will be studied in next Section \ref{Resonance}.

In any case, the perturbative decay rates normally act as a lower bound on the true decay rate. The only reasonable ways they could be shut off is (i) if the decay products were fermions, then there is the issue of Pauli blocking \cite{Cohen:1986ct}. However, this is irrelevant here since our decay products are the bosons $\phi$ themselves. Or (ii) if the star carried sufficient binding energy to make the process kinematically impossible. This is not an issue for dilute stars which carry very small binding energy, while highly compact stars will be addressed in Section \ref{Binding}. 

These perturbative rates can be readily calculated. The boson star (at least away from very high compactness) can be viewed as a condensate of non-relativistic particles. Their large de Broglie wavelengths mean that they overlap with one another and can therefore annihilate through the contact interactions of the potential. This leads to the star emitting energy in the form of pairs of scalar particles. 
As shown in Ref.~\cite{Hertzberg:2010yz} the normalized perturbative rates for energy output $\GamD=|dE_*/dt|/E_*$ (where $E_*=M_*$ is the energy of the star) for $3\phi\to2\phi$ and $4\phi\to2\phi$ annihilation processes are given by 
\ba
\GamD(3\phi\rightarrow2\phi)&=&\frac{\lambdatt^3}{(3!)^2 2^4}\frac{\pi^{5/2}k_{\mbox{\tiny{32}}}}{\Gamma(\frac{3}{2})(2\pi)^3\m^{6}}\frac{\int{d^3x \,n_*^3({\bf x})}}{\int{d^3x\,n_*({\bf x})}}\label{GamD32}\\
\GamD(4\phi\rightarrow2\phi)&=&\frac{\lambdaft^4}{(4!)^2 2^5}\frac{\pi^{5/2}k_{\mbox{\tiny{42}}}}{\Gamma(\frac{3}{2})(2\pi)^3\m^{9}}\frac{\int{d^3x \,n_*^4({\bf x})}}{\int{d^3x \,n_*({\bf x})}}\label{GamD42}\,\,\,\,\,\,\,\,\,\,\,
\ea
where $k_{\mbox{\tiny{32}}}\approx\sqrt{5}\,\m/2$ and $k_{\mbox{\tiny{42}}}\approx\sqrt{3}\,\m$ are the momenta of the 2 outgoing particles in each process, respectively. Here $n_*({\bf x})$ is the number density of particles within the boson star, which can be approximated as $n_*({\bf x})=\en_*({\bf x})/\m$, where $\en_*({\bf x})$ is the local energy/mass density. 
Here the coefficients are given by 
$\lambdatt^3\equiv(5\lambda_3(\lambda_3^2+3\lambda_4)/12+\lambda_5)^2\sim\lambda^3$ and 
$\lambdaft^4\equiv(\lambda_4^2-\lambda_6)^2\sim\lambda^4$, where for completeness we have included the sextic term $\lambda_6\phi^6/(6!\,\m^2)$ in eq.~(\ref{GamD42}) (although we did not include it earlier when discussing the profile of the star). We note that for special choices of couplings, such as $\lambda_4^2=\lambda_6$, these rates can be shut off; but this is not the generic situation. Interestingly, this condition $\lambda_4^2=\lambda_6$ occurs when one Taylor expands a cosine potential, which may describe some kinds of axions; however, since it has attractive interactions, it is not of direct relevance here.

\subsection{Application to Boson Stars}

To evaluate these rates we need the star's density $n_*({\bf x})$. It is difficult to provide exact analytical results as one needs to solve nonlinear differential equations to obtain the star's profile $\Phi(r)$; important work includes Refs.~\cite{Chavanis:2011zi,Chavanis:2011zm}. To proceed it is useful to use an approximate form for the shape of a star, which captures the following 2 basic ideas: (i) it is flat near its core, i.e., $\Phi'(0)=0$, and (ii) it falls off exponentially at large distances. For example, as used in Ref.~\cite{Schiappacasse:2017ham}, a useful representation that has these properties and is somewhat accurate (at least for stars that are not too compact) is a sech ansatz
\be
\Phi(r)\approx\sqrt{\frac{3M_*}{\pi^3\m^2 R^3}}\,\,\mbox{sech}(r/R),
\label{sechansatz}\ee
where $R$ is length scale that should be adjusted to minimize the energy to obtain the most accurate solution. The corresponding energy/mass density is $\en_*({\bf x})=\m^2|\Phi|^2$, and the normalization ensures that the mass of the star is $M_*$. This analysis is quite accurate in the non-relativistic regime of low compactness stars, as shown in Ref.~\cite{Schiappacasse:2017ham}. For compact stars, they involve orbital boson speeds that are $\mathcal{O}(1/2)$ that of light. In that regime these estimates are anticipated to still be correct to within a factor of a few, which suffices for our main results. 

Star solutions involve a relationship between the total mass/energy of the star $M_*$ and its physical radius $\Rp$. A precise definition of the physical radius of a star is the region that contains a fixed percentage of the mass of the star. For definiteness we will take this to be the radius that encloses 90\% of the mass. This turns out to be related to the scale that appears in the argument of the above sech profile by $\Rp\approx d\,R$, with $d\approx2.8$ for the sech profile. However, an $\mathcal{O}(1)$ change in this definition will not be important for our main results. 

The radius-mass relationship for the condensate arises from minimizing the energy from kinetic energy (pressure), repulsive self-interaction, and gravitation. It can be shown this leads to \cite{Chavanis:2011zi,Chavanis:2011zm,Schiappacasse:2017ham}
\beq
\tilde{R} = {a+\sqrt{a^2+3bc\tilde{M}_*^2}\over b\tilde{M}_*}
\label{radiusmass}\eeq
where $a,b,c$ are numerical constants that depend on the choice of ansatz. For example, for the above sech ansatz they are: $a=(12+\pi^2)/(6\pi^2)$, $b=6(12\zeta(3)-\pi^2)/\pi^4$,\,$c=(\pi^2-6)/(8\pi^5)$. Here we are using dimensionless variables  $\tilde{R}\equiv \m^2\,R/(\sqrt{\lambdaeff}\,\mpl)$ and $\tilde{M}_*\equiv\sqrt{\lambdaeff}\,M_*/\mpl$. Note that at large $\tilde{M}_*\gg1$ the radius approaches a constant $\tilde{R}\to\sqrt{3c/b}$, giving a star size of
\be
\Rp\to{d\sqrt{3c\,\lambdaeff}\,\mpl\over\sqrt{b}\,\m^2}\,\,\,\,\,\,\,\,\,\,(M_*\gg\mpl/\sqrt{\lambdaeff})\label{Rasmp}
\ee
This is as expected from the scalings of the full Einstein-Klein-Gordon equations in the previous section. 

Let us focus on this asymptotic regime of the more massive stars, since they are of most interest astrophysically. We can readily carry out the above integrals to obtain the decay rates as
\ba
\GamD(3\phi\to2\phi) & = &\pref_3{\lambdatt^3\, \m^5\,M_*^2\over\lambdaeff^3\,\mpl^6}\label{gam3as}\\
\GamD(4\phi\to2\phi) & = &\pref_4{\lambdaft^4\, \m^7\, M_*^3\over\lambda_4^{9/2}\mpl^9}\label{gam4as}
\ea
where $\pref_3,\,\pref_4$ are $\mathcal{O}(1)$ prefactors. The sech ansatz estimates their values to be $\pref_3\approx 0.04,\,\pref_4\approx 0.05$. Note that in the denominator of the $\GamD(4\phi\to2\phi)$ result we have replaced $\lambdaeff\to\lambda_4$ since it is only relevant if the cubic term $\sim\lambda_3\m\phi^3$ is negligible. 

Recall that for glueballs we anticipate $\lambda_n\sim(4\pi/N)^{n-2}$. So $\lambdatt^3/\lambdaeff^3=\mathcal{O}(1)$ in the expression for $\GamD(3\phi\to2\phi)$, making it essentially independent of the size of the gauge group. While $\lambdaft^4/\lambda_4^{9/2}\sim1/\sqrt{\lambda}\sim N/(4\pi)$ in the expression for $\GamD(4\phi\to2\phi)$, making it increase linearly with the size of the gauge group (if other parameters are kept fixed).

\section{Astrophysical Bounds}\label{Bounds}

If the above annihilation rates (\ref{gam3as},\,\ref{gam4as}) are much bigger than the current Hubble rate $H_0$ of the universe, then the stars are unlikely to be cosmologically relevant. As an example, consider the case with $\lambda=\mathcal{O}(1)$ and $\m\sim 0.1$\,GeV, giving rise to $\Mmax$ of the order of 10s of solar masses. Then, in order for the decay rates eqs.~(\ref{gam3as},\ref{gam4as}) to be smaller than today's Hubble rate $H_0\approx 10^{-33}$\,eV, we obtain the following bound on the mass of the star: $M_*\lesssim 10^{-18}\,\Ms$ if $3\phi\to2\phi$ processes are active and  $M_*\lesssim 10^{-11}\,\Ms$ if $4\phi\to2\phi$ processes are active. Such stars have very low compactness and are nowhere close to being relevant to the LIGO/LISA bands. This already undermines the claims of Refs.~\cite{Soni:2016gzf,Soni:2016yes,daRocha:2017cxu,Soni:2017nlm}, as well as other work on real scalars, including Ref.~\cite{Eby:2015hsq}, since the heavy stars of these analyses would be too short lived. 

\begin{figure}[h!]
    \centering
    \includegraphics[width=1.02\columnwidth,height=8.2cm]{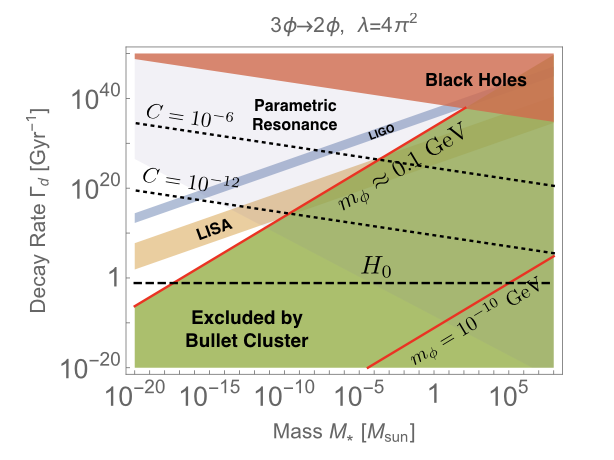}\\
    \includegraphics[width=1.02\columnwidth,height=8.2cm]{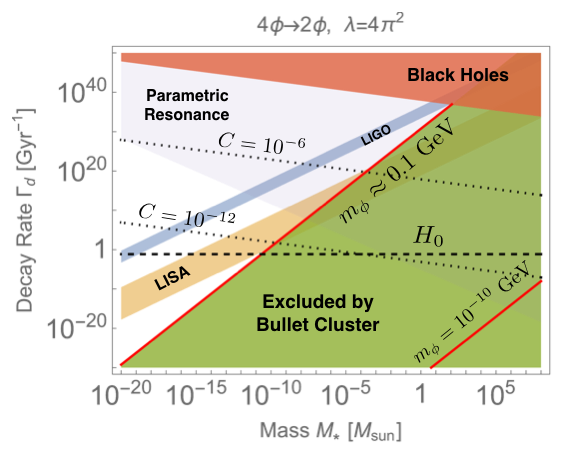}
    \caption{{\em Fixing coupling $\lambda$ to its maximum value}.
Decay rates $\GamD$ (in units Gyr$^{-1}$) of boson stars as a function of the star's mass $M_*$ (units $\Ms$) for different choices of particle mass $\m$ (solid red) and compactness $\C$ (dotted grey).
Here we have imposed the coupling is $\lambda=(2\pi)^2$, which is on the order of the maximum allowed by unitarity (appropriate for small gauge groups).
The black dashed line is the present Hubble rate $H_0$.
In the upper right orange region the stars would collapse to black holes. 
The lower right green shaded region is excluded by bullet cluster bounds.
The yellow band is LISA, and the blue band is LIGO, which is only appreciable at the top right where the stars are compact. 
The light blue shaded region indicates where parametric resonance may occur according to a simple analysis.
Upper panel: $3\phi\to2\phi$ processes when there are odd powers of $\phi$ in $V$.
Lower panel: $4\phi\to2\phi$ processes when there are only even powers of $\phi$ in $V$.}
    \label{RatesFixedLambda}
\end{figure}

\begin{figure}[h!]
    \centering
    \includegraphics[width=1.02\columnwidth,height=8.2cm]{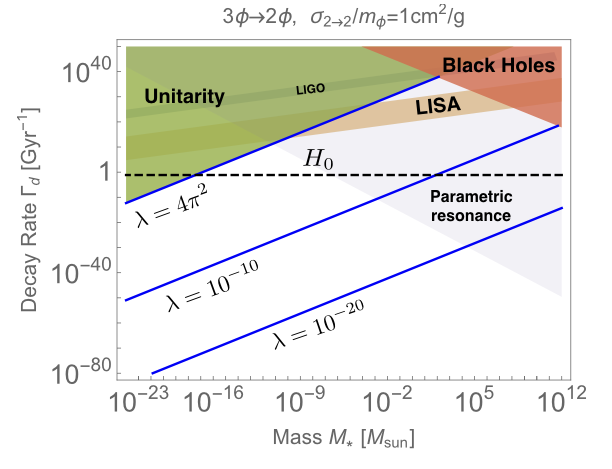}\\
    \includegraphics[width=1.02\columnwidth,height=8.2cm]{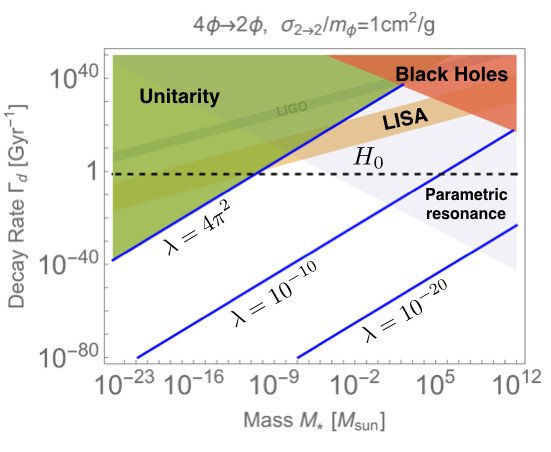}
    \caption{{\em Fixing scattering to core-cusp preferred value}.
Decay rates $\GamD$ (in units Gyr$^{-1}$) of boson stars as a function of the star's mass $M_*$ (units $\Ms$) for different choices of coupling $\lambda=(4\pi/N)^2$ (solid blue). 
Here we have imposed the scattering cross section is $\sigma_{2\to2}/\m=1\,\mbox{cm}^2/\mbox{g}$ to possibly explain the cores of galaxies.
The black dashed line is the present Hubble rate $H_0$.
In the upper right orange region the stars would collapse to black holes. 
In the upper left green there are no stars, since once would need couplings $\lambda\gg 1$ violating unitarity. 
The yellow band is LISA, and the blue band is LIGO, which is only appreciable at the top right where the stars are compact. 
The light blue shaded region indicates where parametric resonance may occur according to a simple analysis.
Upper panel: $3\phi\to2\phi$ processes when there are odd powers of $\phi$ in $V$.
Lower panel: $4\phi\to2\phi$ processes when there are only even powers of $\phi$ in $V$.}
    \label{Rates}
\end{figure}

\begin{figure}
\vspace{0.45cm}
    \centering
    \includegraphics[width=\columnwidth,height=8.4cm]{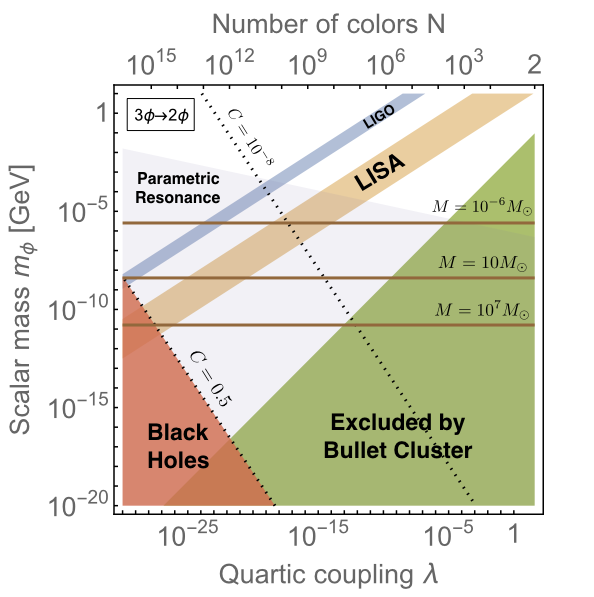}\\
    \includegraphics[width=\columnwidth,height=8.4cm]{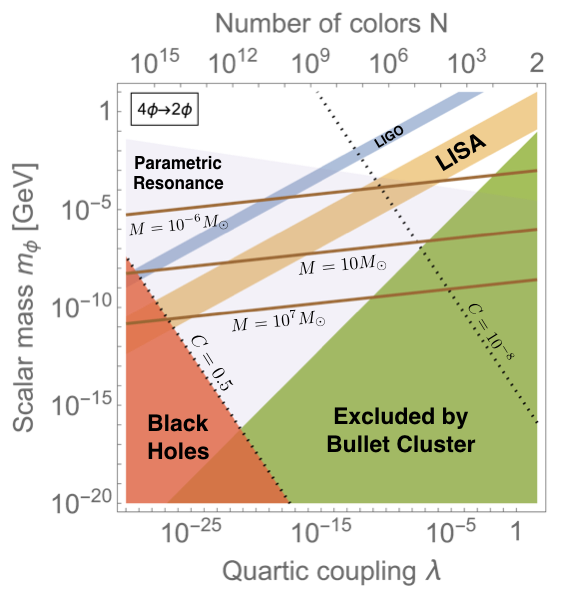}
    \caption{{\em Fixing perturbative decay rate to Hubble and exploring $\{\m,\,\lambda\}$ parameter space}.
    Parameter space of solutions after imposing the decay rate is equal to the current Hubble rate $\GamD=H_0$.
    The solid brown lines are contours of fixed star mass $M_*$ and the dotted black are contours of fixed compactness $\C$. The lower right green shaded region is excluded by bullet cluster bounds. The lower left shaded orange region would have compactness so high the stars would collapse to black holes. 
 The yellow band is LISA, and the blue band is LIGO, which is only appreciable at the lower left where the stars are compact. 
The light blue shaded region indicates where parametric resonance may occur according to a simple analysis.
Upper panel: $3\phi\to2\phi$ processes when there are odd powers of $\phi$ in $V$.
Lower panel: $4\phi\to2\phi$ processes when there are only even powers of $\phi$ in $V$.}
 \label{ParticleMassCoupling}
\end{figure}

\begin{figure}[h!]
\vspace{0.45cm}
    \centering
    \includegraphics[width=\columnwidth,height=8.4cm]{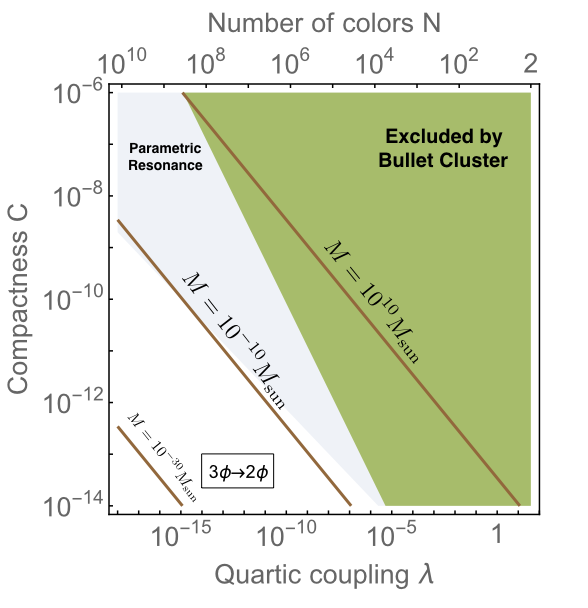}\\
    \includegraphics[width=1.01\columnwidth,height=8.4cm]{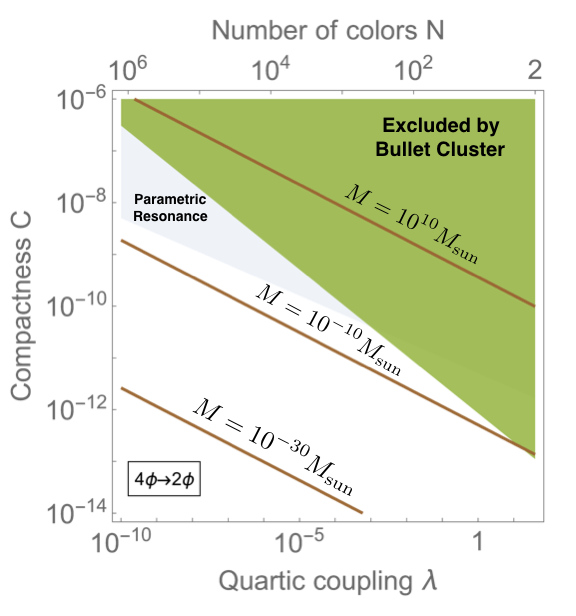}
    \caption{{\em Fixing perturbative decay rate to Hubble and exploring $\{\C,\,\lambda\}$ parameter space}.
    Parameter space of solutions after imposing the decay rate is equal to the current Hubble rate $\GamD=H_0$.
  The solid brown lines are contours of fixed star mass $M_*$. 
    The upper right green shaded region is excluded by bullet cluster bounds. 
   In this figure we have not indicated the LISA or LIGO bands or the black hole regime since these are only important at high compactness, which is far off the top of the plot. 
The light blue shaded region indicates where parametric resonance may occur according to a simple analysis.
Upper panel: $3\phi\to2\phi$ processes when there are odd powers of $\phi$ in $V$.
Lower panel: $4\phi\to2\phi$ processes when there are only even powers of $\phi$ in $V$.}
    \label{CompactnessCoupling}
\end{figure}

A much more general exploration of the constrained parameter space is provided in the Figs.~\ref{RatesFixedLambda},\ref{Rates},\ref{ParticleMassCoupling},\ref{CompactnessCoupling}, where we have included detailed information in their captions.
The decay rates $\GamD$ are plotted versus star mass $M_*$ in Figs.~\ref{RatesFixedLambda},\ref{Rates}, with the coupling $\lambda=(2\pi)^2$ fixed in Fig.~\ref{RatesFixedLambda}, and the scattering cross-section fixed in Fig.~\ref{Rates} (see next subsection for explanation). Then in Figs.~\ref{ParticleMassCoupling},\ref{CompactnessCoupling} we fix the decay rate to be today's Hubble rate $\GamD=H_0$, with contours of fixed mass $M_*$ indicated; these provided upper bounds on the allowed mass for the star to live to present time. We also indicated the compactness (see upcoming subsection for explanation).

In the remainder of this section and the next section we would like to provide more details on the ingredients that have gone into these figures.

\subsection{Scattering in Galaxies}

Apart from the boson stars, one expects there to be a large collection of diffuse $\phi$ particles acting as a form of dark matter. Due to the above self-interactions they will undergo scattering in the galaxy. The $2\phi\to2\phi$ scattering cross section for non-relativistic particles is readily obtained as
\be
\sigma_{2\to2} = {\lambdaeff^2\over 128\pi\m^2}
\ee

If the scalar particles $\phi$  make up a significant fraction, or all, of the dark matter then there are bounds on this scattering. Some of the best bounds come from observations of collisions of galaxies, such as the bullet cluster, which are essentially consistent with non-interacting dark matter. This imposes an observational {\em upper} bound on the scattering cross section of \cite{Markevitch:2003at,Harvey:2015hha}
\be
{\sigma_{2\to2}\over\m}\lesssim \,\bc\,\mbox{cm}^2/g
\ee
where $\bc$ is argued to be $\mathcal{O}(1)$, depending on the analysis. For concreteness, we will take the bound with $\bc=1$ in this work. In Figs.~\ref{RatesFixedLambda},\ref{ParticleMassCoupling},\ref{CompactnessCoupling}, we have imposed this bound, which rules out the green region. We add that if the scalar particles $\phi$ only make up a tiny fraction of the dark matter (yet there is still enough to provide some boson stars), then these bounds are significantly weakened; in this case the green region can largely be ignored.

On the other hand, it has been argued that the cores of galaxies are explained precisely by the presence of such scattering \cite{Spergel:1999mh}. To do so requires a {\em lower} bound on the cross section of $\sigma_{2\to2}/\m\gtrsim \cc\,\mbox{cm}^2/g$, where $\cc$ has also argued to be $\mathcal{O}(1)$. Altogether this provides some motivation to consider the case of $\sigma_{2\to2}/\m\sim \mbox{cm}^2/g$, although it is still controversial. (See Ref.~\cite{Deng:2018jjz} for a critical examination of using ultralight scalars to solve this problem.)  We have fixed the scattering to be $\sigma_{2\to2}/\m\sim \mbox{cm}^2/g$ in Fig.~\ref{Rates}. In that plot we are showing decay rate versus star mass $M_*$ for different choices of the coupling $\lambda$. In the upper left region one would have $\lambda\gg1$, which is forbidden by unitarity. For concreteness we take $\lambda=(2\pi)^2$ as the upper value (although one could argue for even smaller values to be safer from unitarity considerations).

\subsection{Compactness}\label{Compact}

Of particular interest for the production of gravitational waves is the compactness $\C$ of a star. We shall define this as the ratio of the corresponding Schwarzschild 
radius $\Rs=2G M_*$ and the physical radius of the star $\Rp$
\be
\C\equiv{\Rs\over \Rp}\label{Compactdef}
\ee
If the compactness is $\mathcal{O}(1)$, then it is undergoing strong gravity in its vicinity. When $M_*\to\Mmax$, as defined in Section \ref{LCR}, then we are indeed in this regime. The compactness there is roughly $\C\sim 1/2$ and mergers can produce gravitational waves with significant amplitudes. Of course the compactness parameter cannot be larger than 1 or the system would have collapsed to a black hole. In Figs.~\ref{RatesFixedLambda},\ref{Rates},\ref{ParticleMassCoupling} we have indicated this in the orange region (In Fig.~\ref{CompactnessCoupling} this occurs at much higher values of $\C$ than those displayed).
On the other hand, the compactness can be small $\C\ll1$, describing the dilute boson star. In this regime the gravitational wave signal from mergers is expected to be suppressed.

We can consider a family of solutions that exist at some compactness $\C$. For example, one may imagine that the stars will continue to accrete until they have achieved their maximum compactness $\C\sim1/2$. For any particle mass $\m$ we can consider this possibility. So we use eqs.~(\ref{Rasmp},\,\ref{Compactdef}), eliminate $\m$ in favor of $\C$, and insert into the decay rate formulas to obtain
\ba
\GamD(3\phi\to2\phi) & = &\prefC_3{\lambdatt^3\,\mpl^{3/2}\C^{5/2}\over \lambdaeff^{7/4} \,\sqrt{M}_*}\label{gam3asC}\\
\GamD(4\phi\to2\phi) & = &\prefC_4{\lambdaft^4\,\mpl^{3/2}\C^{7/2}\over \lambda_4^{11/4} \,\sqrt{M}_*}\label{gam4asC}
\ea
where $\prefC_3,\,\prefC_4$ are $\mathcal{O}(1)$ prefactors. The sech ansatz estimates their values to be $\prefC_3\approx 0.005,\,\prefC_4\approx 0.0001$. 
Contours of fixed compactness are provided in Fig.~\ref{RatesFixedLambda} for fixed coupling and in Fig.~\ref{ParticleMassCoupling} for fixed decay rate.

\subsection{Implications for Gravitational Waves}

Importantly, by fixing the decay rate to be Hubble, we plot contours of maximum allowed star mass $M_*$ in the $\{\C,\,\lambda\}$-plane in Fig.~\ref{CompactnessCoupling}. This shows that for reasonable couplings, and to obey the bullet cluster bound, the mass and compactness need to be small for longevity. This has significant implications for gravitational waves from possible mergers of these stars. It implies that one is well outside of both the LIGO and LISA bands. Gravitational wave detection requires $\mathcal{O}(10-10^6)$ solar mass objects with high compactness. However these results indicate that the decay rates are too fast to achieve this. For completeness, we have indicated both the LIGO and LISA bands in Figs.~\ref{RatesFixedLambda},\ref{Rates},\ref{ParticleMassCoupling}. In Figs.~\ref{RatesFixedLambda},\ref{Rates} the signals are only appreciable in the upper right hand region, which involves rapid decay. While in Fig.~\ref{ParticleMassCoupling}, where the decay is fixed to Hubble, the signal is only appreciable in the lower left hand region, involving extremely tiny self couplings, and may be unlikely to persist due to the phenomenon of parametric resonance, which we turn to now.

\section{Parametric Resonance}\label{Resonance}

In addition to the quantum mechanical perturbative decays discussed above, one can also enter a regime in which the coherently oscillating boson star condensate drives parametric resonance of its own field fluctuations. Since the stars of interest are wide compared to the inverse mass of the particle ($\Rp\sim\sqrt{g}/\m\gg1/\m$) they are susceptible to parametric resonance. This would represent a type of instability against linear perturbations; differing views on this have been expressed in the literature \cite{Jetzer:1992np,Clayton:1998zza}. 
As mentioned earlier, the criteria for this to occur is
\be
\mu\,\Rp>1\label{inequality}
\ee
where $\mu$ is the maximum exponential growth rate (``Floquet exponent") within the homogeneous background approximation. The intuition behind this is that when this inequality is not satisfied, the produced particles escape the condensate before Bose-Einstein statistics are effective. In any case, it was earlier established in Ref.~\cite{Hertzberg:2010yz} (also see earlier work in the context of axion-photons in Refs.~\cite{Tkachev:1986tr,Tkachev:1987cd,Levkov:2020txo}), so we shall not repeat the derivation here. Hence we do not need to consider the full complications of expanding around the inhomogeneous star background, we can focus on the corresponding homogenous configuration with amplitude matching the star's core amplitude, obtain $\mu$, and check on this inequality.

Let us perturb around the background as
\be
\phi({\bf x},t)=\phi_0(t)+\delta\phi({\bf x},t)
\ee
Here one should, in principle, also allow for fluctuations in the metric. However, the metric fluctuations are primarily only important to describe {\em long} wavelength perturbations around the homogeneous background. This leads very importantly to the collapse of homogeneous structure and is the source of structure formation, and ultimately to the formation of boson stars, etc. However, what we are interested in is to imagine a star has formed and we are only interested in these annihilation processes inside its core. These are {\em particle number changing} processes and are mediated by self-interactions of the potential, and are not mediated by gravity (one could consider resonance into gravitons but this is normally highly suppressed). Hence we can focus on flat space fluctuations for the purpose of this discussion. The perturbed equation of motion is
\be
\ddot{\delta\phi}-\nabla^2\delta\phi +\m^2\delta\phi+V_I''(\phi_0(t))\delta\phi=0\label{linearEOM}
\ee
where $V_I$ is the interaction potential. The advantage of studying this homogenous condensate is that it can be readily diagonalized by passing to Fourier space $\delta\phi\to\delta\phi_k$ and $-\nabla^2\delta\phi\to k^2\delta\phi_k$. This makes eq.~(\ref{linearEOM}) a form of Hill's equation since it is a linear differential equation with a periodically changing prefactor $V_I''(\phi_0(t))$.

\subsection{Floquet Exponents}

Let us begin with the case when both odd and even powers in the potential are included. The most obvious version is the cubic term $\lambda_3\m\phi^3/3!$ and quartic $\lambda_4\phi^4/4!$. However to obtain the relevant resonance is slightly complicated (as can be appreciated by drawing all the corresponding Feynman diagrams). So for the sake of simplicity, let us focus on an interaction provided by $V_I=\lambda_5\,\phi^5/(5!\,m)$, which provides $3\phi\to2\phi$ with no intermediate propagators. Although, one should also anticipate the presence of cubic terms, but the final result will have a similar scaling (assuming $\lambda_n\sim\lambda^{(n-2)/2}$).

To first approximation, the oscillating condensate is mainly driven by the mass term. So the oscillations of the background are
\be
\phi_0(t)\approx\phia\cos(\omegah t)
\ee
where $\omegah\approx\m$ and $\phia$ is the amplitude of oscillations. By inserting this back into eq.~(\ref{linearEOM}) one obtains the interaction term as a pair of harmonics due to the factor
\be
V_I''(\phi_0(t))\propto\phi_0^3(t) = {\phia^3\over4}\left(3\cos(\omegah t)+\cos(3\omegah t)\right)\label{phi3}
\ee
One can then expand $\delta\phi$ in harmonics too. One can have resonance from long wavelength perturbations, which can be driven by the leading harmonic term. However, as mentioned above this is not important for us. We know that this only leads to the destabilization of the homogeneous condensate towards a boson star etc. Instead we are interested in the possible resonance at the higher harmonic $3\omegah\approx3\m$. This is potentially resonant for $\delta\phi_k\propto e^{i 3\omegah/2}$, since when we insert this into the equation of motion, the driving term will have a frequency that matches the input frequency. In turn this matches the natural frequency for $\omega_k\equiv\sqrt{k^2+m^2}=3\omegah/2\approx 3\m/2$, which means $k\approx k_{\mbox{\tiny{32}}}\approx\sqrt{5}\,\m/2$ the outgoing wave number mentioned earlier in our perturbative analysis in eq.~(\ref{GamD32}).  

Hence in the vicinity of this resonance of interest, we can write the equation of motion as
\be
\ddot{\delta\phi}_k+\omega_k^2\delta\phi_k+{\lambda_5\phia^3\over4\!\cdot\!3!\,\m}\cos(3\omegah t)\delta\phi_k=\mbox{n.r}\label{linearEOM5s}
\ee
where \mbox{``n.r"} refers to the non-resonant term from the $\cos(\omegah t)$ in eq.~(\ref{phi3}). Ignoring the non-resonant piece, this is a form of the Mathieu equation
\be
{d^2\over d\tau^2}\delta\phi_k+(A_k+2B\cos(2\tau))=0
\ee
which is known to possess exponential growth in some band of wavenumbers. Here we can identify $A_k=\omega_k^2/(3\omegah/2)^2$, $B=\lambda_5\phia^3/(8\cdot3!\,\m)/(3\omegah/2)^2$. For small amplitudes, the Floquet exponent is known to be \cite{Hertzberg:2014jza}
\be
\mu_k={3\omegah\over4}\sqrt{B^2-(A_k-1)^2}
\ee
As we scan over different $k$-values this is clearly maximal when $A_k=1$, i.e., for $\omega_k=3\omegah/2$ as expected. This gives the maximum Floquet exponent of (using $\omegah\approx\m$)
\be
\mu_{3\to2}
={|\lambda_5|\,\phia^3\over144\,\m^2} \to {\lambdatt^{3/2}\,\phia^3\over144\,\m^2}
\ee
where we have indicated in the final step that when we include the cubic and quartic couplings, we can generalize the result to
$|\lambda_5|\to|\lambda_5+5\lambda_3(\lambda_3^2+3\lambda_4)/12|=\lambdatt^{3/2}$, since we know this arises from the scattering amplitudes.

If there are no odd powers of $\phi$ in the potential, we can still have parametric resonance from the quartic term $V_I=\lambda_4\phi^4/4!$. In this case the analysis is rather more complicated. This can be seen by the fact that there are now several Feynman diagrams associated with the $4\phi\to2\phi$ process. The full details of this analysis was carried out by some of us in Ref.~\cite{Hertzberg:2014jza}.
A simpler calculation arises from a sextic term $V_I=\lambda_6\phi^6/(6!\,\m^2)$ because it occurs through a process without any intermediate processes. For this we can again follow again the above analysis, this time expanding $V_I''(\phi_0(t))\propto\phi_0^4(t)$ in harmonics and focussing on the $\cos(4\omegah t)$ term. The result is
\be
\mu_{4\to2} = {|\lambda_6|\,\phia^4\over1536\,\m^3} \to {\lambdaft^2\,\phia^4\over1536\,\m^3}\label{mu42final}
\ee
where we have indicated that we can generalize this to $\lambda_6\to|\lambda_6-\lambda_4^2|=\lambdaft^2$, since we know this is the relevant contribution from both processes.

\subsection{Application to Boson Stars}

We now would like to apply these results to the boson star. Firstly we need to relate $\phia$ to the properties of the star. The idea of the criteria $\mu\,\Rp>1$ for resonance is that one replaces the the amplitude of the homogenous configuration $\phia$ by the corresponding amplitude at the core of the star, i.e., 
\be
\phia\to\sqrt{2}\,\Phi_0=f\sqrt{2M_*\over\m^2\,R_*^3}\label{phiareplace}
\ee 
where $f$ is yet another $\mathcal{O}(1)$ prefactor; in the sech ansatz it is given by $f=\sqrt{3\,d^3/\pi^3}\approx1.46$. Hence the product of interest is found to be
\ba
\mu_{3\to2}\,\Rp&=&\prefmu_3\,{\lambdatt^{3/2}\,\m^2\,M_*^{3/2}\over\lambdaeff^{7/4}\,\mpl^{7/2}}\label{muR32}\\
\mu_{4\to2}\,\Rp&=&\prefmu_4\,{\lambdaft^2\,\m^3\,M_*^2\over\lambda_4^{5/2}\,\mpl^5}\label{muR42}
\ea
where $\prefmu_3=f^3b^{7/4}/(108\sqrt{2}\,3^{3/4}c^{7/4}d^{7/2})$
and $\prefmu_4=f^4b^{5/2}/(3456\sqrt{3}\,d^5c^{5/2})$. 
In the sech ansatz they are given by $\prefmu_3\approx2.1$
and $\prefmu_4\approx1.8$.

The region in which the inequality for parametric resonance is obeyed $\mu\,\Rp>1$ is indicated by the light blue region in Figs.~\ref{RatesFixedLambda},\ref{Rates},\ref{ParticleMassCoupling},\ref{CompactnessCoupling}. We see that it can constrain the allowed parameter space considerably. However, its scaling is different to the perturbative decays, so it is often complementary.

If we now eliminate the particle mass $\m$ to rewrite $\mu\,\Rp$ in terms of the compactness $\C$, as we did in Section \ref{Compact}, we obtain
\ba
\mu_{3\to2}\,\Rp&=&\prefmuC_3\,{\lambdatt^{3/2}\,\C\sqrt{M_*}\over\lambdaeff^{5/4}\sqrt{\mpl}}\label{muR32C}\\
\mu_{4\to2}\,\Rp&=&\prefmuC_4\,{\lambdaft^2\,\C^{3/2}\sqrt{M_*}\over\lambda_4^{7/4}\sqrt{\mpl}}\label{muR42C}
\ea
where $\prefmuC_3,\,\prefmuC_4$ are $\mathcal{O}(1)$ prefactors. The sech ansatz estimates their values to be $\prefmuC_3\approx0.4,\,\prefmuC_4\approx0.1$.  This result suggests something important: By imposing the bound on mass $M_*\gg\mpl/\sqrt{\lambdaeff}$, so that we are in the self-interaction supported regime (as opposed to the quantum pressure supported regime) and using $\lambda_n\sim\lambda^{(n-2)/2}$, we have $\mu_{3\to2}\,\Rp\gg \C$ and $\mu_{4\to2}\,\Rp\gg\C^{3/2}$. This means that if one considers stars of large compactness, say $\C\sim 1/2$, then the inequality eq.~(\ref{inequality}) is satisfied and parametric resonance is expected to occur. This suggests very compact stars that could undergo mergers and be relevant to LIGO/LISA are likely to be quite unstable to parametric resonance. However, since we have not solved for the time dependence fully we can not be certain of this conclusion. This leads us to search for additional clues, as we do in the next section.

\section{Binding Energy}\label{Binding}

The results of the previous sections indicate that any massive compact stars will decay rapidly. However, one might consider the possibility that the boson stars carry so much binding energy that these processes are shut off. The above analyses were done reliably in the weak gravitational field regime, where the binding energy is small and is unlikely to be able to shut off these processes. However, in the strong gravity regime this becomes at least conceivable.

To examine this fully for a real scalar, we would really need to solve the full set of time dependents equations by expanding in a tower of harmonics, as we described in Section \ref{SHA}. However, as we also discussed there, this is a rather difficult task and will be left for future work.

\begin{figure}[t]
    \centering
    \includegraphics[width=\columnwidth]{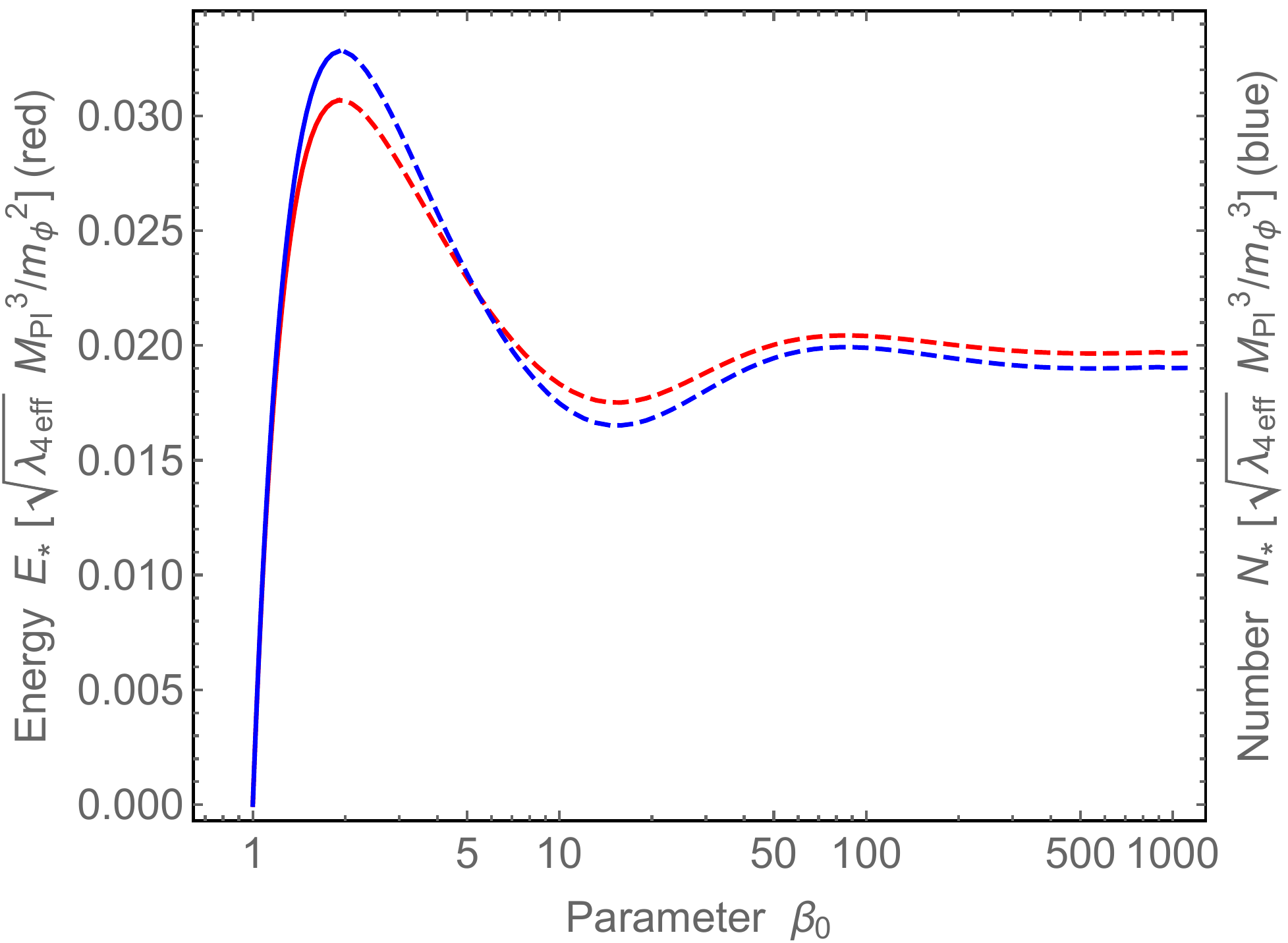}
    \caption{Boson star's energy/mass (red) and number (blue) as a function of parameter $\bp_0$ within the large $\F$ approximation. (The left solid curves are the solutions of most interest in this work, with the maximum value $\Mmax\approx0.03\sqrt{\lambdaeff}\,\mpl^3/\m^2$. The right dashed curves might suffer from other sorts of instabilities.)}
    \label{MassNumberBeta}
\end{figure}

For now we will simply take our clues from the much simpler case of a complex scalar field theory. To define this, we can return to our starting action eqs.~(\ref{Saction}), replace the real scalar $\phi$ by a complex scalar $\psi$, and endow the theory with a global $U(1)$ symmetry. To be concrete, the kinetic term is now $\Delta\mathcal{L}=-|\partial\psi|^2/2$ and we choose the potential now as $V=\m^2|\psi|^2/2+\lambda_4|\psi|^4/16$. The $U(1)$ symmetry ensures there is a conserved current \cite{Friedberg:1986tp}
\be 
J^\mu = i\sqrt{|g|}\,g^{\mu\nu}(\psi^*\partial_\nu\psi-\psi\,\partial_\nu\psi^*)/2
\ee
with a corresponding conserved particle number
\be 
N_*=\int d^3x\,J^0(x,t) 
\ee
One can now rigorously define the boson star as a state that minimizes the energy subject to the constraint of fixed particle number. It is simple to show that such solutions have exactly the simple time dependence
\be
\psi(r,t)=\Phi(r)\,e^{-i\omega t}
\ee
(or $\psi(r,t)=\Phi(r)\,e^{+i\omega t}$ for a star of anti-particles). 
The conserved particle number is
\be 
N_*=4\,\pi\,\omega\!\int_0^\infty dr\,r^2\sqrt{\frac{A(r)}{B(r)}}\,\Phi^2(r)
\ee
Note that for the complex scalar, the metric of a single boson star is exactly static, so we could write $\bar{A}=A$, $\bar{B}=B$ here.
By solving the large $\F$ Einstein-Klein-Gordon equations, we can determine the star's energy/mass and number for a family of solutions; this is given in Fig.~\ref{MassNumberBeta}. The left hand solid curves are the usual solutions that we have focussed on in this work; they are stable for a complex scalar. The right hand dashed curves may have instabilities of a variety we have not discussed in this work; we leave their analysis for future work.

\begin{figure}[t!]
    \centering
    \includegraphics[width=0.92\columnwidth]{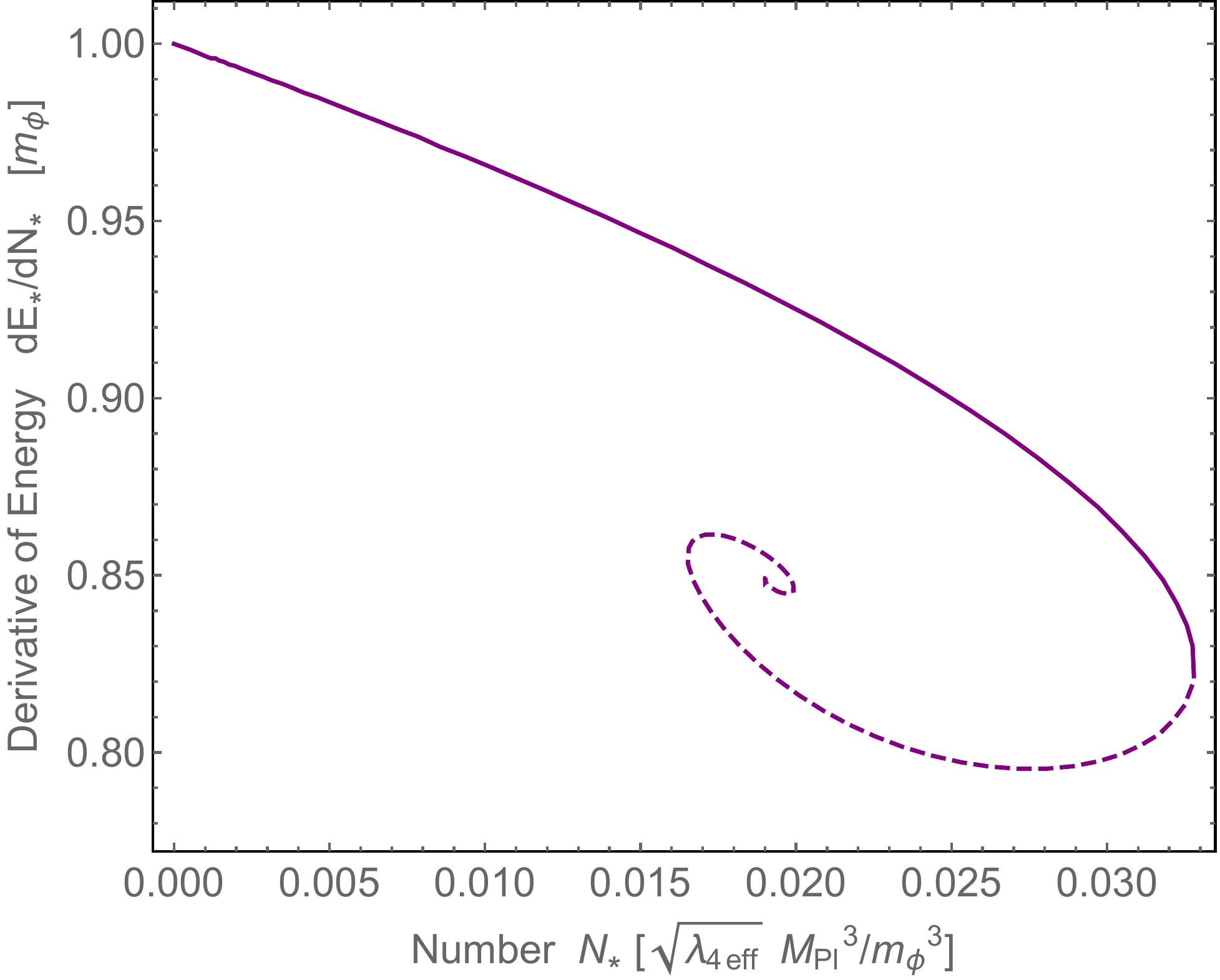}
    \caption{Change in boson star's energy with respect to particle number. (The upper left solid curve are the regular solutions of most interest in this work. The lower right dashed curve might suffer from other sorts of instabilities.)}
    \label{BindingEnergy}
\end{figure}

Now we would like to infer any ramifications for the case of a real scalar. We can be concrete about this in the following way: Suppose we have the above complex scalar with a $U(1)$ symmetry, and now we introduce small $U(1)$ breaking terms, which can in principle mediate particle number changing processes. In order for gravity to prevent an annihilation process of the form $n\,\phi\to2\phi$, the change in energy with respect to particle number would need to be less than $2\m/n$. So to kinematically forbid $3\rightarrow2$ annihilations one would need $dE_*/dN_* < 2\m/3$, while to kinematically forbid $4\phi\to2\phi$ annihilations one would need $dE_*/dN_*<\m/2$. In Fig.~\ref{BindingEnergy} we show our results for $dE_*/dN_*$ as a function of particle number. We see that for all stars up to the maximum value allowed (solid branch), we find
\be
{d E_*\over dN_*}>0.82\,\m
\ee
(and $d E_*/dN_*>0.79\,\m$ including the dashed branch).
So, while there can be an appreciable amount of binding energy for the most massive stars, it does not appear to be quite enough to prevent annihilations. So although we have not performed the full analysis for the real scalar, this provides some circumstantial evidence that binding energy, even for the most compact stars, will not be enough to prevent the decay processes computed in the previous sections.

\section{Classical ``Annihilation" in Cores}\label{Classical}

So far we have considered quantum annihilation in the core of a star. What we mean by this is the expansion given in eq.~(\ref{heisenberg}), i.e., we take the star solution $\phi_*(r,t)$, which we take to be exactly periodic and localized, and we add the inevitable fluctuations required by quantum mechanics by the operator $\delta\hat{\phi}({\bf x},t)$. This is very important in the perturbative regime, since it is this bath of quantum fluctuations that get driven through {\em forced} resonance to provide a stream of outgoing particles, with corresponding annihilation rates $\GamD(3\phi\to2\phi)$ or $\GamD(4\phi\to2\phi)$. As a matter of principle it is also important for the parametric resonance, even though the growth (Floquet) rates can be computed classically. That is because one also needs to perturb the solution in order to see the growth. Such perturbations can very easily happen from imperfect initial conditions around the star solutions, and so could be seen classically, but if one starts exactly on the star solution, then it is quantum fluctuations that are important. 

However, in addition to this is the fact that the boson star solution $\phi_*(r,t)$ is in general not expected to be an exact solution of the classical equations of motion. This is because it is generally very difficult to have any exact periodic solutions of non-linear partial differential equations. One of the only known counter examples is the sine-Gordon breather in $1+1$-dimensions, but that is not our focus here.
So if one expands the solution in harmonics at small amplitudes, one anticipates that the expansion is only an asymptotic series, missing exponentially small terms. These corrections are sometimes referred to as part of a ``hyperasymptotic series" \cite{Boyd2}. Such series have been addressed in the literature for quite sometime \cite{Segur:1987mg}, so we will only report on some of its basic features here. 

In order to address this, one can write the full {\em classical} boson star solution as
\be
\phi(r,t) = \sum_{n=1}^\N\Phi_n(r)\cos(\omega_nt)+\cs(r,t)
\ee
where the sum is the exactly periodic piece, summed up to $\N$ terms in the harmonic expansion, and
 $\cs(r,t)$ is the inevitable correction that survives at that order. 
If we then insert this into the full equations of motion and expand to linear order in $\cs$ (since $\cs$ is expected to be very small), the structure of the resulting equation is
\be
\square\cs-\m^2\cs = J(r,t)\label{cs}
\ee
where $J(r,t)$ arises from the star solution which is never exact at any order $\N$ in the expansion. Due to the presence of nonlinear terms, the driving term $J$ is will take the form
\be
J(r,t)= j(r)\cos(\omega_2 t)+\ldots\label{source}
\ee
where $\omega_2$ is the frequency of the leading harmonic that is generated by the nonlinear equations of motion from the fundamental $\omega_1$. For a theory with odd powers of $\phi$ in the potential, $\omega_2=2\omega_1$, while for a theory with only even powers of $\phi$ in the potential, $\omega_2=3\omega_1$. Here $j(r)$ is a function of radius that is determined by the star solutions $\Phi_n(r)$. 
Eqs.~(\ref{cs},\ref{source}) imply that the boson star acts as a coherent oscillating source that is generating its own classical scalar radiation $\cs$. The relevant solution of eq.~(\ref{cs}) is the particular solution, which is readily obtained in Fourier space as $\cs=\int d^4k/(2\pi)^4\, J(k,\omega)e^{i({\bf k}\cdot{\bf x}-\omega t)}/(-\omega^2+k^2)$. In the far distance regime, well outside the star, the tail of this radiation takes the form
\be
\cs(r,t)\sim{\cos(k_2r+\gamma)\over r}\cos(\omega_2 t)\,j(k_2)\label{csrad}
\ee
where $k_2$ is the wavenumber of the outgoing on-shell radiation with $\omega_2=\sqrt{k_2^2+\m^2}$ and $j(k_2)$ is the Fourier transform of $j(r)$ evaluated at $k_2$. 

As demonstrated in Ref.~\cite{Hertzberg:2010yz} the value $j(k_2)$ is on the order of the star's solution itself $j(k_2)\sim\m^2\,\phi_*(k_2)$ (evaluated at, say, $t=0$); this is reasonable since it is the star solution that provides the source for its own radiation. Then using eq.~(\ref{csrad}) the corresponding power output is therefore
\be
\Big{|}{dE\over dt}\Big{|} \sim |\m^3\,\phi_*(k_2)|^2\sim|\m^3\,\Phi(k_2)|^2
\ee
where in the last step we are using the single harmonic approximation, which was studied in detail in Section \ref{SHA}, and we are using $\omega\sim\m$ here. 

For the sake of concrete analytical results, let us use the sech ansatz of eq.~(\ref{sechansatz}), which is most trustworthy for dilute stars. We can readily obtain its Fourier transform as 
\be
\Phi(k)=\sqrt{3\pi^3\,M_* R\over k^2\m^2}\,\mbox{tanh}\!\left(\pi k R\over2\right)\mbox{sech}\!\left(\pi k R\over2\right)
\ee
We need to evaluate this at the on-shell wavenumber $k=k_2=\sqrt{\omega_2^2-\m^2}$. For stars of low compactness, we have $\omega_1\approx\m$. So then can write $k_2=\kappa_2\,\m$, where $\kappa_2=\sqrt{3}$ when there are odd powers of $\phi$ in the potential, and $\kappa_2=\sqrt{8}$ when there are only even powers of $\phi$ in the potential. The stars of interest are much wider than the inverse mass of the field, i.e., $R\,\m\gg 1$. So we are deep into the tails of both the tanh and sech functions above. In this regime, we can approximate tanh$(\pi k_2 R/2)\approx1$ and sech$(\pi k_2R/2)\approx 2\,\exp(-\pi \kappa_2\m R/2)$. The corresponding decay rate due to this classical radiation output is $\GamC=|dE_*/dt|/E_*$ ($E_*=M_*$ is energy of the star). This gives (dropping $\mathcal{O}(1)$ factors)
\be
\GamC \sim \Rp\,\m^2\,\exp(-c\,\kappa_2\,\m\,\Rp)\label{GamCstar}
\ee
where $c$ is an $\mathcal{O}(1)$ number; which has the value $c=\pi/d\approx1.1$ within the sech ansatz. However, we expect there to be an $\mathcal{O}(1)$ correction to $c$ from the real solution, so its specific value here is not the focus. This leads to the expectation of exponential suppression. 

However, we note that the above Fourier transform is really only valid in a small amplitude expansion, ensuring that the star is wide and dilute and enters a scaling regime controlled by one scale $R$ (we shall return to study this case in detail in Section \ref{PressureStars}). In the more massive branch of solutions of most interest in this paper, the star actually has new features; including the somewhat more vertical shape at the edge of the star seen in Fig.~\ref{StarSolution} when $\F$ is large. Hence this simple estimate of the Fourier transform is only trustworthy for dilute stars, and may be less applicable for dense stars with another scale entering the analysis. Nevertheless, we only wish to comment on a very basic qualitative feature of the solution: In particular, for stars of large radius $\m\,\Rp\gg 1$, one can still anticipate the Fourier transform is somewhat small at $k=k_2=\kappa_2\,\m\sim\m$. Indeed we know that for boson stars supported by repulsive self-interactions, the radius is on the order the coupling introduced earlier $\Rp\,\m\sim\sqrt{\F}\sim\sqrt{\lambdaeff}\,\mpl/\m$, which is assumed to be large so far in our work. 

Hence, the classical radiation for $M_*\ll \Mmax$ boson stars is expected to be small. There may be a large change in this conclusion for somewhat compact stars, wherein more bosons are circulating in a relativistic fashion. In this case the Fourier transform could have appreciable support at the relevant wavenumber $k_2=\kappa_2\,\m\sim\m$, and the classical radiation could be significant. But from our analyses in earlier sections, quite compact stars are already anticipated to have significant perturbative, and possible parametric, decays anyhow.

\subsection{Comparison to Literature}

The above analysis involves the star itself driving its own classical radiation. In a sense this can be viewed as a kind of $2\to1$ process, in the case in which there are odd powers of $\phi$ in the potential, or a a kind of $3\to1$ process, in the case in which there are only even powers of $\phi$ in the potential. While this does not immediately seem to be compatible with momentum conservation, if we recall that the outgoing waves are spherically symmetric, there is no real problem.

In Ref.~\cite{Eby:2015hyx}, a similar type of $3\to1$ process was considered. There the focus was on axions in which there are expected to be only even powers of $\phi$ in the potential. In that work they considered the limit in which gravity is decoupled. In this regime, one is not in fact even studying boson stars. Instead in this regime these are oscillons (``axitons" in Ref.~\cite{Kolb:1993hw}); scalar field solutions in which the {\em attractive} self-interactions ($\lambda_4<0$) are balanced by the field's gradient pressure (``quantum pressure"). (For connections to dark matter; see Ref.~\cite{Olle:2019kbo}). Even though this is a different regime to what we are mainly focussing on in this work, we can nevertheless compare the basic form of the radiation formulae.

In this case there is a well defined small amplitude expansion (although it suffers from a collapse instability in 3+1-dimensions), and one can compute the radiation output from oscillons (for example, see Refs.~\cite{Fodor:2009kf,Hertzberg:2010yz}). Again defining an instantaneous rate $\GamC\equiv|dE_*/dt|/E_*$ for convenience, one finds
\be
\GamC = K\,{\m\over\varepsilon}\,\exp(-\bar{c}\,\kappa_2/\varepsilon)\label{GamClassical}
\ee
where the coefficient of the exponential can be reliably found to be $\bar{c}\approx 1.2$. Here $\varepsilon\ll1$ is a small expansion parameter; it is related to the shift in frequency of the fundamental by $\omega_1^2=\m^2(1-\varepsilon^2)$. In this small amplitude regime, it also sets the size of the oscillon to be $\m\,\Rp\sim1/\varepsilon$. Hence the scaling here is similar to the case above in eq.~(\ref{GamCstar}), albeit a difference is that $\Rp$ is fixed by the gravitational mass-radius relation eq.~(\ref{radiusmass}) for the stars of interest, while it is not fixed to this value for the oscillon. The coefficient $K$ depends on the specific form of the potential. It is normally $K=\mathcal{O}(1)$ for a generic even potential. This should include the QCD axion, whose potential is expected to be complicated due to contributions from many instantons \cite{diCortona:2015ldu}.  However, if one considers the special case of a single cosine potential (from a single instanton which may be relevant to other kinds of axions), it is suppressed by cancellations (one can see similar behavior in our earlier results eqs.~(\ref{GamD42},\ref{mu42final}) which had prefactors proportional to $\lambda_{42}^2=|\lambda_4^2-\lambda_6|$ vanishing for the expansion of the cosine); in this case the value is only $K=\mathcal{O}(\varepsilon^2)$ \cite{Fodor:2009kf}. 

On the other hand, in Ref.~\cite{Eby:2015hyx} they studied decay within the single cosine potential, and the corresponding $3\to1$ process was claimed to be quantum mechanical. They considered transition matrix elements between the $|N_*\rangle$ and $|N_*-3\rangle$ axion quantum states. This led them to obtain the scaling for the rate of $3\to1$ annihilations as
\be
\Gamma_3\sim {\m\over\lambda\,\varepsilon^2}\,\exp(-\bar{c}\,\kappa_2/\varepsilon)
\label{Eby}\ee
To compare notation to Ref.~\cite{Eby:2015hyx} replace $\varepsilon\to\Delta$ and $\lambda\to\m^2/f_a^2$, where $f_a$ is the PQ scale of the axion. We note that the exponent is the same as our results above.
This is the rate at which 3 particles are converted into 1 particle. 
Then to obtain the decay rate of the entire condensate, one divides by the number of particles in the condensate (or say, half the number) in order to obtain the characteristic time for an appreciable fraction (say half) of the condensate to radiate away. The number of particles in the oscillon condensate, in this small amplitude expansion, can be readily shown to be $N_*\sim 1/(\lambda\,\varepsilon)$. So by forming $\Gamma/N_*$, one obtains a form of the scaling in eq.~(\ref{GamClassical}) with $K=\mathcal{O}(1)$. While this would be correct for a generic even potential, this misses the important cancellations that take place for a pure cosine potential, in which one should actually obtain $K=\mathcal{O}(\varepsilon^2)$. Since modern treatments of the QCD axion indicate that it is not in fact too close to a pure cosine \cite{diCortona:2015ldu}, then $K=\mathcal{O}(1)$ should in fact be correct for the QCD axion. However, the pure cosine was in fact the subject of Refs.~\cite{Eby:2015hyx,Eby:2017azn}, so their scaling was partially incorrect.

We note that if one analyzes eq.~(\ref{Eby}) as a decay rate, one finds something worthy of note: if we were to reinstate factors of $\hbar$, starting with the classical field theory, one can readily show that it is proportional to $1/\hbar$. This means its classical field theory limit is badly behaved. Instead, by noting that this is not the physical rate of decay of the macroscopic condensate, but instead we need to divide by a factor on the order $\sim N_*\sim 1(\hbar\,\lambda\,\varepsilon)$, we then obtain the scaling of eq.~(\ref{GamClassical}) (with $K=\mathcal{O}(1)$) which is independent of $\hbar$ and is indeed classical. Hence even though Ref.~\cite{Eby:2015hyx} derived this decay rate through a quantum mechanical analysis, the corresponding physical decay rate is ultimately purely a property of the classical field theory. Instead the actual quantum rates appear very differently, and are {\em not} exponentially suppressed; see eqs.~(\ref{GamD32},\,\ref{GamD42}).

\section{Quantum Pressure Supported Stars}\label{PressureStars}

For completeness, let us also briefly discuss more familiar boson stars: those that are not supported by repulsive $\lambda\phi^4$ interactions, but instead are supported by the field's gradient pressure (which is often referred to as ``quantum pressure", as it is quantum mechanical from the particle point of view, since it originates from the de Broglie wavelength of the bosons). In this regime the couplings can in fact be taken to zero $\lambda\to0$ and the stars still persist. It can be readily shown that in the $\lambda\to0$ limit the maximum mass of such stars is on the order (e.g., see Ref.~\cite{Helfer:2016ljl})
\be
\Mmax \sim {\mpl^2\over\m}
\label{MaxDilute}\ee
Which is a very different scaling to the maximum mass in the self-interaction supported regime of $\Mmax\sim\sqrt{\lambda}\,\mpl^3/\m^2$ which has been the focus of this paper up until now. In particular, this now requires extremely small particle masses $\m$ for the maximum mass to be astrophysically significant.

The regime of quantum pressure supported stars is when the condition in eq.~(\ref{Rasmp}) is no longer satisfied, i.e., consider now the opposite regime $M_*\ll\mpl/\sqrt{\lambdaeff}$. In this regime the radius of the star is inverse proportional to the star's mass (see eq.~(\ref{radiusmass}))
\be
\Rp\approx{2\,a\,d\,\mpl^2\over b\,\m^2\,M_*}\,\,\,\,\,\,\,\,\,\,(M_*\ll\mpl/\sqrt{\lambdaeff})
\ee
This is sometimes referred to as the ``dilute boson star". However, we note that so long as $\lambdaeff$ is extremely small, namely $\lambdaeff\lesssim\m^2/\mpl^2$, this can be still rather compact in this regime. 

\subsection{Perturbative Decays}

By inserting this scaling of the radius into the perturbative decay rate eqs.~(\ref{GamD32},\ref{GamD42}), we obtain
\ba
\GamD(3\phi\rightarrow2\phi)&=&\prefD_3\,{\lambdatt^3\,\m^5\,M_*^8\over\mpl^{12}}
\label{GamD32dilute}\\
\GamD(4\phi\rightarrow2\phi)&=&\prefD_4\,{\lambdaft^4\,\m^7\,M_*^{12}\over\mpl^{18}}
\label{GamD42dilute}
\ea
where the coefficients $\prefD_3,\,\prefD_4$ are prefactors that turn out to be rather small. In the sech ansatz they are given by $\prefD_3\approx 6\times 10^{-10}$ and $\prefD_4\approx 9\times 10^{-14}$. 

It is convenient to introduce the parameterization $\lambda \sim\F\,\m^2/\mpl^2$. We can then readily bound the decay rates. In the regime $\F\gtrsim1$ it is useful to use the fact that to stay within this regime we need $M_*\ll\mpl/\sqrt{\lambdaeff}$. While if $\F\lesssim1$ it is convenient to use the fact that there is a maximum mass in this regime $M_*\lesssim\Mmax\sim\mpl^2/\m$. By imposing these inequalities on the decay rate we obtain the bounds
\ba
\GamD(3\phi\rightarrow2\phi)&\lesssim&\prefD_3\,{\m^3\over\mpl^2}\,\mbox{Min}\!\left[\F^{-1},\,\F^3\right]
\label{GamD32bound}\\
\GamD(4\phi\rightarrow2\phi)&\lesssim&\prefD_4\,{\m^3\over\mpl^2}\,\mbox{Min}\!\left[\F^{-2},\,\F^4\right]
\label{GamD42bound}
\ea
Hence these rates are extremely small for any interesting values of $\m$, since we need extremely small $\m$ to have massive stars by eq.~(\ref{MaxDilute}). 
This implies that perturbative decays are negligible in this regime. 

\subsection{Parametric Resonance}

We can also evaluate the parametric resonance rates by inserting into eqs.~(\ref{muR32},\ref{muR42}) and we obtain
\ba
\mu_{3\to2}\,\Rp&=&\prefmuD_3\,{\lambdatt^{3/2}\,\m^2\,M_*^{5}\over\mpl^7}\label{muR32Dilute}\\
\mu_{4\to2}\,\Rp&=&\prefmuD_4\,{\lambdaft^2\,\m^3\,M_*^7\over\mpl^{10}}\label{muR42Dilute}
\ea
where the coefficients $\prefmuD_3,\,\prefmuD_4$ are prefactors that again turn out to be somewhat small. In the sech ansatz they are given by $\prefmuD_3\approx 6\times 10^{-5}$ and $\prefmuD_4\approx 5\times 10^{-7}$.

We can again form inequalities as we did earlier, by making use of $M_*\ll\mpl/\sqrt{\lambdaeff}$ for $\F\gtrsim1$ and $M_*\lesssim\Mmax\sim\mpl^2/\m$ for $\F\lesssim1$. Together this gives the bounds
\ba
\mu_{3\to2}\,\Rp&\lesssim&\prefmuD_3\,\mbox{Min}\!\left[\F^{-1},\,\F^{3/2}\right]\label{muR32DiluteBd}\\
\mu_{4\to2}\,\Rp&\lesssim&\prefmuD_4\,\mbox{Min}\!\left[\F^{-3/2},\,\F^{2}\right]\label{muR42DiluteBd}
\ea
Hence we always have $\mu\,\Rp\ll1$ and the inequality for resonance is never satisfied.
This implies that parametric resonance is also negligible in this regime. 

Altogether then, bosons stars formed out of real scalars that are supported by quantum pressure are robust again decays, while bosons stars supported by self-interactions that we analyzed in earlier sections are often not.

\section{Summary and Outlook}\label{Conclusions}

In this work we have analyzed the stability of boson stars built out of dark matter scalars due to repulsive self-interactions. Since the mass of such stars can be as large as $\Mmax\sim\sqrt{\lambda}\,\mpl^3/\m^2$ they are potentially astrophysically relevant. For glueballs, or any other motivated scalar particle, one may consider $\m\sim 0.1$\,GeV and $\lambda=\mathcal{O}(1)$ and obtain stars of several solar masses at high compactness. In the literature it was suggested that this may potentially produce interesting gravitational wave signatures if mergers occur. However, since glueball stars are built out of real scalars in the effective theory, rather than a complex scalar with a global $U(1)$ symmetry, they have no conserved particle number. We computed the perturbative annihilation rates $\GamD$ in eqs.~(\ref{gam3as},\ref{gam4as}) in terms of fundamental parameters $\m$, $\lambda_n$, and the star's mass $M_*$, or if we eliminate $\m$ in favor of the star's compactness $\C$ in eqs.~(\ref{gam3asC},\ref{gam4asC}). For $\m\sim0.1$\,GeV and $\lambda=\mathcal{O}(1)$ these decays rates are found to be much quicker than the current Hubble rate, unless the stars are many orders of magnitude lighter than a solar mass. Such tiny stars would not be astrophysically relevant and would have tiny compactness. 

We explored the full parameter space of possibilities in Section \ref{Bounds}, finding that there is essentially no reasonable parameters for which the star is massive, compact, and long lived. We further investigated the possibility of parametric resonance of fluctuations in Section \ref{Resonance}, with the relevant dimensionless parameter being $\mu\,\Rp$ to indicate whether the Bose-Einstein statistics are effective or not, given in eqs.~(\ref{muR32},\ref{muR42}) or eqs.~(\ref{muR32C},\ref{muR42C}). We found that the resonance becomes the dominant mechanism for stability at smaller couplings and can considerably constrain the parameter space further.

An important caveat is that all of our analysis was done with a simple starting point for the star, in which we describe it within the time averaged single harmonic approximation. To then ascertain decay rates, we then perturb around this to obtain quantum decay rates by the Heisenberg equation of motion due to forced resonance or exponential growth rates due to parametric resonance. This strategy should be valid for low compactness stars, which are essentially non-relativistic and possess a reasonable single harmonic approximation. However, this could be less accurate at high compactness, where many harmonics are expected to play a role, and this simple analysis could conceivably miss out on some of the physics. We did study the binding energy of the star in Section \ref{Binding} within this simplified treatment, finding that it does not appear to be enough to prevent decays, but a full treatment would be preferable.

Important future work is therefore to included the full time dependence, which can be written as a collection of harmonics and then to solve a system of ODEs for the coefficient functions of radius. Alternatively, one could run full simulations, which has the advantage of not only capturing all harmonics, but identifying instabilities readily. This is at least true for the instabilities associated with parametric resonance, while quantum radiation can be much harder to see in a classical simulation, as it requires seeding the fluctuations with a bath of zero point fluctuations, which can be difficult to track reliably in a finite simulation.

We also examined the classical decay rate in Section \ref{Classical}, which is ordinarily much smaller because the star's are wide. This means that, except perhaps for somewhat compact stars, their Fourier transform is typically small at the relevant resonant wavenumber, so they do not efficiently emit their own classical radiation. Conversely, the quantum radiation is set by $\hbar$ and is not suppressed in most regimes of interest. We compared the basic structure of the classical rate to existing claims in the literature \cite{Eby:2015hyx}, regarding $3\to1$ processes for dilute axions, in which they studied the limit in which gravity is decoupled. Here the condensate is in fact a type of oscillon (or ``axiton") and is held together by attractive self-interactions. The work of Ref.~\cite{Eby:2015hyx} quite rightly claimed the process was exponentially suppressed at small amplitudes, but claimed that this effect is quantum mechanical. They obtained the quantum rate $\Gamma_3$ to transition from the $|N_*\rangle$ to $|N_*-3\rangle$ axion states, which indeed depends on $\hbar$. Instead, we explained that once one divides the rate by the number of particles in the condensate to obtain the physical decay rate, rather than merely the rate for any particle to meet others, one obtains a final result which is independent of $\hbar$, and is therefore classical, as expected. (Moreover, their pre-factor missed cancellations that occur for a pure cosine.) Since the classical rate is exponentially suppressed (at small amplitude) it is normally very small compared to the truly quantum decay rates that we computed here; see eqs.~(\ref{GamD32},\,\ref{GamD42}).

For completeness, we then considered the more regular branch of solutions of boson stars in Section \ref{PressureStars}, namely those that are supported against gravity by ``quantum pressure", rather than repulsive self-interactions. In this regime, we found that both the perturbative and parametric resonance rates are suppressed. This is not too surprising, since the appearance of this branch only emerges in the limit in which self-interactions are small, but it is those same terms that are trying to drive the number changing processes. In any case, these stars are therefore the most robust against decays. On the other hand, they have the parametrically much smaller maximum mass of $\Mmax\sim\mpl^2/\m$, which can only be astrophysically significant for extremely small $\m$, requiring extremely small $\lambda\lesssim\m^2/\mpl^2$ to avoid the self-interaction corrections. 

Other future directions are to return to consider the complex scalar $\psi$ with a global $U(1)$ symmetry \cite{Colpi:1986ye}. This immediately prevents the decay processes studied here, since it has a conserved particle number. However, there are suggestions from quantum gravity that all global symmetries should be explicitly broken. This could introduce terms like $\Delta\mathcal{L}\sim\psi^{4+n}/\mpl^n+c.c$, and leads to the expectation that there may be particle number changing processes allowed in the star. If $n=1$ this is roughly equivalent to replacing $\lambdatt^{3/2}\to\m/\mpl$ in our above decay rates, or if $n=2$ this is roughly equivalent to replacing $\lambdaft^2\to\m^2/\mpl^2$. This can be such a large suppression that it significantly opens up much more parameter space where the stars are stable; however, we leave a full investigation for future work. 

Furthermore, one can consider other types of novel compact stars that may be allowed by other interesting dynamics in dark sectors.  Alternatively, one may focus on other kinds of observational consequences that may arise from dilute stars, including fast radio bursts \cite{Tkachev:2014dpa} or pulsar timing \cite{Siegel:2007fz}.

\section*{Acknowledgments}
M.~P.~H and J.~Y acknowledge support from the Tufts Global Research Assistant Program.
M.~P.~H. is supported in part by National Science Foundation Grant No. PHY-2013953.


\end{document}